\documentclass[draft]{agujournal2019}
\usepackage{url} 
\usepackage{lineno}
\usepackage[inline]{trackchanges} 
\usepackage{soul}
\usepackage{amsmath}
\usepackage{amssymb}
\usepackage{xcolor}

\definecolor{uc-pc}{rgb}{0.58892, 0.288894, 0.213625}
\definecolor{uc-p1}{rgb}{0.999873, 0.680072, 0.67995}
\definecolor{uc-p2}{rgb}{0.792935, 0.48208, 0.443269}
\definecolor{uc-p3}{rgb}{0.381463, 0.125606, 0.043693}
\definecolor{uc-p4}{rgb}{0.21447, 0.06271, 0.001412}
\definecolor{uc-mc}{rgb}{0.197878, 0.501258, 0.64903}
\definecolor{uc-m1}{rgb}{0.621082, 0.690182, 0.999507}
\definecolor{uc-m2}{rgb}{0.385296, 0.651674, 0.879368}
\definecolor{uc-m3}{rgb}{0.12409, 0.321259, 0.415071}
\definecolor{uc-m4}{rgb}{0.068618, 0.15304, 0.195818}

\nolinenumbers
\draftfalse

\journalname{AGU Advances}

\begin{document}

%
%


\title{How does ice shell geometry shape ocean dynamics on icy moons?}

%
%




\authors{Yixiao Zhang\affil{1},
    Wanying Kang\affil{1},
    and John Marshall\affil{1}
}

\affiliation{1}{Earth, Atmospheric, and Planetary Sciences, Massachusetts
Institute of Technology, Cambridge, MA 02139 US}




\correspondingauthor{Yixiao Zhang}{yixiaoz@mit.edu}


\begin{keypoints}
\item Ice shell topography drives ocean circulation by inducing lateral temperature gradients through suppression of the freezing point of water.
\item The resulting thermal wind supports baroclinic eddies that are very efficient in meridional heat transport.
\item Sloped topography weakens or strengthens heat transport depending on
salinity, which sets the sign of the thermal expansion coefficient.

\end{keypoints}

%
%

%
%


\begin{abstract}

A poleward-thinning ice shell can drive circulation in the subsurface oceans of
icy moons by imposing a meridional temperature gradient--colder at the equator
than the pole--through the freezing point suppression due to pressure. This
temperature gradient sets a buoyancy gradient, whose sign depends on the thermal
expansion coefficient determined by ocean salinity. Together with vertical
mixing, this buoyancy forcing shapes key oceanic features, including zonal
currents in thermal wind balance, baroclinic instability of those currents,
meridional heat transport by eddies, and vertical stratification.
We use high-resolution numerical simulations to explore how variations in ice
shell thickness affect these processes. Our simulations span a wide range of
topographic slopes, pole-to-equator temperature differences, and vertical mixing
strengths, for both fresh and salty oceans. We find that baroclinic eddies
dominate large-scale circulation and meridional heat transport, consistent with
studies assuming a flat ice-ocean interface. However, sloped topography
introduces new effects: when lighter water overlies denser water along the
slope, circulation weakens as a stratified layer thickens beneath the poles.
Conversely, when denser water lies beneath the poles, circulation strengthens as
topography increases the available potential energy.
We develop a scaling framework that predicts heat transport and stratification
across all simulations.
Applying this framework to Enceladus, Europa, and Titan, we infer ocean heat
fluxes, stratification, and tidal energy dissipation and showing large-scale
circulation constrains tidal heating and links future observations of ice
thickness and rotation to subsurface ocean dynamics.
\end{abstract}

\section*{Plain Language Summary}
Icy moons such as Enceladus and Europa host oceans beneath thick ice shells that
are typically thinner at the poles than at the equator, likely due to stronger
tidal heating near the poles. This topography drives a meridional temperature
gradient through pressure-dependent freezing point suppression, creating a
buoyancy gradient that powers ocean circulation. The direction of this
circulation depends on the thermal expansion coefficient of water, which changes
sign between fresh and salty oceans near the freezing point. Using
high-resolution numerical simulations, we investigate how variations in ice
shell shape affect ocean stratification, zonal currents, and meridional heat
transport. We find that baroclinic eddies--swirling motions arising from
instabilities of zonal flows—dominate meridional heat transport. Sloped
topography strongly modulates circulation: when denser water lies beneath the
poles, circulation strengthens; when lighter water lies beneath the poles, it
weakens. We develop scaling laws that predict ocean heat transport and
stratification across a wide parameter range. Applying them to Enceladus,
Europa, and Titan, we find that ocean circulation can limit the amount of tidal
energy stored in the ocean. These results suggest that surface
observations--such as ice thickness and nonsynchronous rotation--could reveal
signatures of subsurface ocean dynamics, with implications for habitability.

\section{Introduction}
    \label{sec:introduction}

Icy moons with subsurface oceans are compelling targets for the search for life
beyond Earth. These worlds, including Europa, Enceladus, and Titan, harbor vast
liquid water oceans beneath their icy shells
\cite{Carr-Belton-Chapman-et-al-1998:evidence,
Khurana-Kivelson-Stevenson-et-al-1998:induced,
Pappalardo-Belton-Breneman-et-al-1999:does,
Kivelson-Khurana-Russell-et-al-2000:galileo,
Thomas-Tajeddine-Tiscareno-et-al-2016:enceladus},
providing potentially habitable
environments \cite{
Taubner-Pappenreiter-Zwicker-et-al-2018:biological,
Glein-Waite-2020:carbonate,
Chyba-2000:energy,
Russell-Murray-Hand-2017:possible}. Understanding the patterns of ocean circulation and the driving mechanisms in these bodies is crucial to assessing their
habitability and interpreting observations of their surface features and
potential biosignatures.

The poleward thinning of the ice shell, likely ubiquitous on icy moons due to the poleward amplified tidal heating pattern
\cite{Ojakangas-Stevenson-1989:thermal, beuthe2018enceladus},
can drive ocean circulation by inducing under-ice temperature and consequent buoyancy
gradients. As the thickness of the ice shell decreases poleward, the pressure at the
ice-ocean interface drops, causing the freezing point to rise. This temperature, when transmitted into the interior by vertical mixing, can induce zonal currents which are prone to baroclinic instability.
The resulting eddies transport heat down the gradient, from the polar
regions to the equator \cite{zhang2024ocean}. This circulation and the resultant
heat transport have been investigated by \citeA{zhu2017influence, lobo2021pole,
kang2022different, zhang2024ocean} using idealized models, theoretical scaling
laws, and numerical 3D simulations. 
The scaling law derived in \citeA{kang2022different, zhang2024ocean} neglected
the actual topography, the fact that the cold water under equatorial ice
is lower in elevation than the warm water under the polar ice shell. 

The difference in the elevations of buoyant and dense water sources strongly
influences the overturning strength and the stratification of the ocean. 
From an energetic perspective, the relative height of the buoyancy source and sink determines whether the gravitational potential energy stored in the ocean will increase or decrease through buoyancy forcing \cite{jansen2022energetics}. Our goal here is to expand on our previous studies \cite{kang2022different,
zhang2024ocean}, and to present a quantitative scaling law that accounts for the tilt of the
ocean top and how it affects meridional heat transport and ocean stratification.
Our scaling framework becomes a powerful tool for linking key components of the
moon system -- including ice geometry
\cite{kang2020spontaneous,kang2022different,kang2022icy}, mean ocean salinity
\cite{zeng2021ocean, kang2022does}, tidal dissipation, the ice shell's energy
balance, and the moon's non-synchronous rotation
\cite{ashkenazy2023non,kang2024nonsynchronous} -- through the lens of large-scale
ocean circulation.

Our paper is set out as follows. Section~\ref{sec:model}
describes the setup of our numerical model and key physical parameters. Section~\ref{sec:solutions} describes representative equilibrium solutions. Section~\ref{sec:scaling} presents our scaling theory, including an analysis of the heat balance and energetics of the eddy field.  Section~\ref{sec:results} tests these scaling predictions against these equilibrium solutions.
Section~\ref{sec:implications} explores
how different parameters affect ocean circulation, and discusses the upper bound on the heat flux. Section~\ref{sec:discussion} discusses the implications for icy moon oceans.

\section{Model Setup}
    \label{sec:model}

\subsection{Under-ice buoyancy forcing}

Poleward-thinning ice shells create an equator-to-pole temperature difference
under the ice, denoted $\Delta T$, with warmer temperatures at the poles and
cooler temperatures at the equator. The upper boundary of an icy moon's
subsurface ocean is defined by the overlying ice, where the temperature equals
the local freezing point, $T_\mathrm{freezing}$. Because $T_\mathrm{freezing}$
depends on pressure, a lateral gradient in ice thickness $\Delta H_i$ induces a
corresponding temperature gradient beneath the ice.

\begin{figure}
    \centering
    \includegraphics[width=\linewidth]{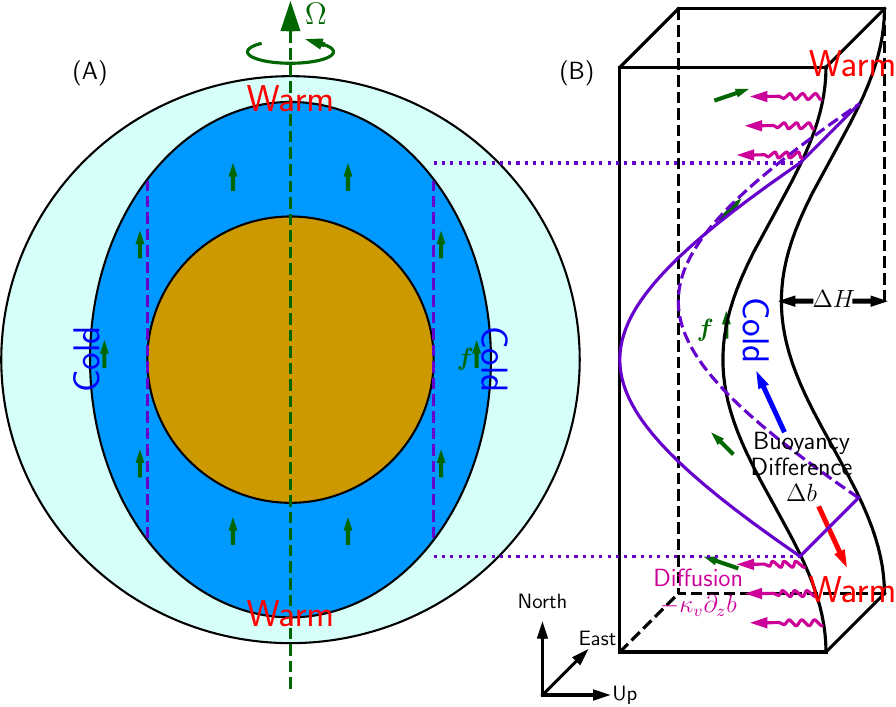}
    \caption{ Setup of our numerical simulations.
        Panel (A) illustrates an idealized icy moon with a poleward-thinning
        ice shell. This ice shell thickness gradient creates a meridional
        temperature gradient at the water-ice interface (``Cold'' and ``Warm''
        in Panel A) due to freezing point suppression by high pressure. The
        resulting temperature contrast $\Delta T$ leads to a buoyancy contrast
        $\Delta b = \alpha g \Delta T$, where $\alpha$ is the thermal expansion
        coefficient and $g$ is the gravitational acceleration. The sign of
        $\Delta b$ depends on the sign of $\alpha$. This buoyancy gradient
        drives ocean circulation, which, under background rotation (indicated by
        arrows in Panel A), gives rise to baroclinic eddies -- referred to as
        ``ocean weather systems'' in \citeA{zhang2024ocean}. Panel (B) depicts
        the numerical setup. Simulations are conducted in a Cartesian coordinate
        system. The poleward-thinning ice shell is represented as a tilted
        upper boundary with an elevation difference $\Delta H$ (see
        Eq.~\ref{eq:z-top}). A buoyancy contrast, consistent with the imposed
        temperature difference, is prescribed as a boundary condition at the
        ocean top (Eq.~\ref{eq:b-top}) and diffuses into the ocean interior via
        vertical diffusion ($-\kappa_v \partial_z b$ in Panel B). The Coriolis
        force is implemented in Cartesian coordinates by conserving the angle
        between the rotation vector $\mathbf{f}$ (green arrows in Panel B) and
        gravity, following \citeA{bire2022exploring, zhang2024ocean}. Parameters
        for each simulation are listed in Table~\ref{tab:simulation}.
    }
    \label{fig:model-setup}
\end{figure}

Assuming hydrostatic equilibrium, the induced temperature difference is:
\begin{linenomath*}\begin{equation}
\Delta T = \rho_w g \Delta H \frac{\mathrm{d}T_\mathrm{freezing}}{\mathrm{d}p},
\label{eq:delta-T}
\end{equation}\end{linenomath*}
where $\rho_w$ is the density of water, $g$ is the gravitational acceleration, and
${\mathrm{d}T_\mathrm{freezing}}/{\mathrm{d}p}$ is the pressure sensitivity of
the freezing point. This sensitivity (approximately $8\times
10^{-8}\,\mathrm{K}\,\mathrm{Pa}^{-1}$) depends only on the latent heat of melting
and the density difference between water and ice, as constrained by the
Clausius-Clapeyron relation. Here, $\Delta H$ is the elevation difference of the
topography of the water-ice interface (Fig.~\ref{fig:model-setup}(B)). Assuming
isostasy \cite{cadek2016enceladus, tajeddine2017true, hemingway2019enceladus,
mckinnon2021new} and neglecting spatial variations in $g$ within the ocean and
ice shell, $\Delta H = (\rho_i/\rho_w)\Delta H_i$, where $\rho_i$ is the
density of ice.

Due to the generally poleward-amplified tidal heating pattern, we focus on cases
with poleward-thinning ice shells, resulting in $\Delta T > 0$--i.e., warmer
poles and cooler equator (Fig.~\ref{fig:model-setup}(A)).

This temperature difference beneath the ice leads to a buoyancy difference,
$\Delta b$, which can drive ocean circulation. We adopt an idealized equation of
state with a constant thermal expansion coefficient $\alpha_T$ in each
simulation and neglect the effects of salinity variations due to ice dynamics
\cite{zhu2017influence, ashkenazy2018dynamics, ashkenazy2021dynamic,
lobo2021pole, kang2022does}.
The buoyancy, defined as $b = -g\,\delta\rho/\rho_w$, where $\delta\rho$ is the
density anomaly, is linearly related to the temperature as:
\begin{linenomath*}\begin{equation}
b = \alpha_T g (T - T_0),
\label{eq:eos}
\end{equation}\end{linenomath*}

with $T_0$ as a reference temperature. The corresponding buoyancy difference is:
\begin{linenomath*}\begin{equation}
\Delta b = \alpha_T g \Delta T,
\end{equation}\end{linenomath*}
where $\alpha_T$ may be positive or negative depending on whether the ocean is
sufficiently salty to offset the anomalous expansion of water \cite{zeng2021ocean,
kang2022does, bire2023divergent}. In this study, we explore both
positive and negative $\Delta b$ values (Table~\ref{tab:simulation}). Since
$\Delta T$ is always positive (reflecting poleward-thinning ice), the sign of
$\Delta b$ is entirely determined by the sign of $\alpha_T$.

\begin{table}
    \caption{Simulation Parameters}
    \centering
    \begin{tabular}{lrrrr}
        \hline
        Group &
        Simulation & $\Delta b$ ($10^{-5}\,\mathrm{m}\,\mathrm{s}^{-2}$)
        & $\Delta H$ (km) & $\kappa_v$ ($10^{-2}\,\mathrm{m}^{2}\,\mathrm{s}^{-1}$) \\
        \hline
        Control ($\Delta b > 0$)
        &ctrl$^+$ \color{uc-pc}{$\blacklozenge$}
        &              1.670 &          6  &          1.0 \\
        \hline
        Varying $\Delta b$
        &b1$^+$   \color{uc-p1}{$\bigstar$}
        & \textit{    0.167} &          6  &          1.0 \\
        &b2$^+$   \color{uc-p2}{$\bigstar$}
        & \textit{    0.501} &          6  &          1.0 \\
        &b3$^+$   \color{uc-p3}{$\bigstar$}
        & \textit{    5.010} &          6  &          1.0 \\
        &b4$^+$   \color{uc-p4}{$\bigstar$}
        & \textit{   16.700} &          6  &          1.0 \\
        \hline
        Varying $\Delta H$
        &h1$^+$   \color{uc-p1}{$\blacksquare$}
        &             1.670  & \textit{ 0} &          1.0 \\
        &h2$^+$   \color{uc-p2}{$\blacksquare$}
        &             1.670  & \textit{ 3} &          1.0 \\
        &h3$^+$   \color{uc-p3}{$\blacksquare$}
        &             1.670  & \textit{12} &          1.0 \\
        &h4$^+$   \color{uc-p4}{$\blacksquare$}
        &             1.670  & \textit{20} &          1.0 \\
        \hline
        Varying $\kappa_v$
        &k1$^+$    \color{uc-p1}{$\blacktriangle$}
        &             1.670  &          6  & \textit{ 0.1} \\
        &k2$^+$    \color{uc-p2}{$\blacktriangle$}
        &             1.670  &          6  & \textit{ 0.3} \\
        &k3$^+$    \color{uc-p3}{$\blacktriangle$}
        &             1.670  &          6  & \textit{ 3.0} \\
        &k4$^+$    \color{uc-p4}{$\blacktriangle$}
        &             1.670  &          6  & \textit{10.0} \\
        \hline
        Control ($\Delta b < 0$)
        &ctrl$^-$  \color{uc-mc}{$\blacklozenge$}
        &            -1.670  &          6  &          1.0 \\
        \hline
        Varying $\Delta b$
        &b1$^-$    \color{uc-m1}{$\bigstar$}
        & \textit{$ -$0.167} &          6  &          1.0 \\
        &b2$^-$    \color{uc-m2}{$\bigstar$}
        & \textit{$ -$0.501} &          6  &          1.0 \\
        &b3$^-$    \color{uc-m3}{$\bigstar$}
        & \textit{$ -$5.010} &          6  &          1.0 \\
        &b4$^-$    \color{uc-m4}{$\bigstar$}
        & \textit{$-$16.700} &          6  &          1.0 \\
        \hline
        Varying $\Delta H$
        &h1$^-$    \color{uc-m1}{$\blacksquare$}
        &         $ -$1.670  & \textit{ 0} &          1.0 \\
        &h2$^-$    \color{uc-m2}{$\blacksquare$}
        &         $ -$1.670  & \textit{ 3} &          1.0 \\
        &h3$^-$    \color{uc-m3}{$\blacksquare$}
        &         $ -$1.670  & \textit{12} &          1.0 \\
        &h4$^-$    \color{uc-m4}{$\blacksquare$}
        &         $ -$1.670  & \textit{20} &          1.0 \\
        \hline
        Varying $\kappa_v$
        &k1$^-$    \color{uc-m1}{$\blacktriangle$}
        &         $ -$1.670  &          6  & \textit{ 0.1} \\
        &k2$^-$    \color{uc-m2}{$\blacktriangle$}
        &         $ -$1.670  &          6  & \textit{ 0.3} \\
        &k3$^-$    \color{uc-m3}{$\blacktriangle$}
        &         $ -$1.670  &          6  & \textit{ 3.0} \\
        &k4$^-$    \color{uc-m4}{$\blacktriangle$}
        &         $ -$1.670  &          6  & \textit{10.0} \\
        \hline
        \multicolumn{5}{p{1.0\linewidth}}{note. 
            In the table $\Delta b$ is the equator-to-pole buoyancy difference
            prescribed at the ocean-ice interface (positive means dense water at
            the equator); $\Delta H$ is the elevation difference of the
            ocean-ice interface from the equator to the pole; $\kappa_v$ is the
            vertical diffusivity of buoyancy. Simulation tags follow the
            following convention: letters indicate the group, with ``ctrl'' for
            the control set, ``b'' for varying $\Delta b$, and ``h'' for varying
            $\Delta H$. The trailing $+$ or $-$ denotes the sign of $\Delta b$:
            $+$ for dense water at the equator and $-$ for light water at the
            equator. A symbol for each simulation is shown next to its tag and
            will be used consistently in the figures. Parameters which differ
            from the control simulations (ctrl$^+$ and ctrl$^-$) are italicized
            for emphasis. A diagram of the model setup is presented in
            Fig.~\ref{fig:model-setup}(B).
        }
    \end{tabular}
    \label{tab:simulation}
\end{table}

\subsection{Governing equations and model details}

We consider ocean circulation driven by the buoyancy gradient along the
water-ice interface, induced by pressure-dependent freezing point suppression
(Fig.~\ref{fig:model-setup}). The essential ingredients of this circulation
include:
1) an equator-to-pole under-ice buoyancy difference $\Delta b$ that drives the
flow, 2) an elevation difference at the water-ice interface between the origins
of dense and buoyant water masses, 3) vertical diffusion that transmits the
upper-boundary buoyancy gradient into the ocean interior, and 4)
the Coriolis force.
As illustrated in Fig.~\ref{fig:model-setup}, the key physical parameters
controlling the ocean dynamics include the equator-to-pole buoyancy contrast
$\Delta b$, the ice thickness variation $\Delta H$, the vertical diffusivity
$\kappa_v$, and the moon's rotation rate $\Omega$, its radius $a$, and mean ocean
depth $D$.

To represent the system in numerical simulations, we follow a similar approach
to \cite{zhang2024ocean} but with a key modification: we account for the actual
ice topography using immersed boundary conditions instead of a flat top
boundary. We use a Cartesian coordinate system configured to represent flow
in a spherical shell, as sketched in Fig.~\ref{fig:model-setup}(B). The extent of
the domain in the eastward $x$ and northward $y$ directions are $[0,
L_x]$ and $[-\pi/2 a, \pi/2 a]$ respectively, where $L_x$ is the zonal
domain size, $a$ is the planetary radius. The elevation of the ocean top
is set to a cosine function:
\begin{linenomath*}\begin{equation}
    z_\mathrm{top} = -\frac{1}{2}\Delta H\cos{\left(2y/a\right)},
    \label{eq:z-top}
\end{equation}\end{linenomath*}
and the ocean bottom is always at $-D$. The
western ($x = 0$) and eastern ($x = L_x$) boundaries are periodic. 

Our governing equation is the rotating, non-hydrostatic equations for a
Boussinesq fluid (referred to as the Boussinesq model):
\begin{linenomath*}\begin{equation}
    \left\{\begin{aligned}
    \partial_t \mathbf{u} + \mathbf{u}\cdot\nabla\mathbf{u}
        + \mathbf{f}\times\mathbf{u}
    &= -\nabla P
        + b \mathbf{e}_z
        + \nu_h\nabla_h^2\mathbf{u} + \nu_v\partial_z^2\mathbf{u}; \\
    \partial_t b + \mathbf{u}\cdot\nabla b &=
        \kappa_h\nabla_h^2 b + \kappa_v \partial_z^2 b; \\
    \nabla\cdot\mathbf{u} &= 0,
    \end{aligned}\right.
    \label{eq:boussinesq}
\end{equation}\end{linenomath*}
where $\mathbf{u} = (u, v, w)$ is the vector form velocity;
$\mathbf{f}$ is the Coriolis parameter; $P$ is the pressure
divided by the reference density, a diagnostic field to
keep $\mathbf{u}$ divergence-free; $b$ is buoyancy; $\nu$
and $\kappa$ represent viscosity and diffusivity, and the subscripts $v$ and $h$
represent the vertical and horizontal components, respectively.
The Boussinesq model (Eq.~\ref{eq:boussinesq}) uses buoyancy $b$ instead of 
temperature $T$ as the prognostic field. To retrieve the temperature field $T$,
one needs to use the equation of state (Eq.~\ref{eq:eos}).

At the top of the ocean, the buoyancy $b$ is set to a cosine profile:
\begin{linenomath*}\begin{equation}
    b_\mathrm{top} = -\frac{1}{2}\Delta b\cos{\left(2y/a\right)}.
    \label{eq:b-top}
\end{equation}\end{linenomath*}
The sign of $\partial_y b_\mathrm{top}$ in the Northern Hemisphere
is the same as $\Delta b$. The bottom, north, and south
boundaries have zero normal flux. A linear frictional stress
of $(-\gamma u, -\gamma v)$ is applied to the velocity components $u$ and $v$ at the top and bottom boundaries.
In this study, $\gamma$ is set to a constant of $1\times
10^{-4}\,\mathrm{m}\,\mathrm{s}^{-1}$ \cite{jansen2022energetics}.

The Coriolis parameter is chosen to mimic that in a spherical shell and is given by:
\begin{linenomath*}\begin{equation}
    \mathbf{f} = 2\Omega \exp\left(\frac{z}{a}\right)
        \left(
                \cos\left(\frac{y}{a}\right)\mathbf{j}
                + \sin\left(\frac{y}{a}\right)\mathbf{k}
        \right)
\end{equation}\end{linenomath*}
This form guaranties that the angle between $\mathbf{g}$ and $\mathbf{f}$ is as
it is on a rotating sphere, and that $\mathbf{f}$ is non-divergent, ensuring
the conservation of potential vorticity in adiabatic inviscid flow
\cite{grimshaw1975note,dellar2011variations}.

We run a set of simulations to test how circulation strength and ocean heat
transport vary with $\Delta b$ (the contrast of equator-to-pole buoyancy),
$\Delta H$ (the difference of ice thickness from equator-to-pole) and $\kappa_v$ (the
vertical diffusivity). The parameter choices are summarized in
Table~\ref{tab:simulation}. In all simulations, $D = 30\,\mathrm{kilometer}$;
$a = 252.3\,\mathrm{kilometer}$;
$L_x = 102\,\mathrm{kilometer}$; the horizontal grid size is
$\Delta x = \Delta y = 1.2\,\mathrm{kilometer}$ and the
vertical grid size $\Delta z = 0.6\,\mathrm{kilometer}$;
$\kappa_h$ and $\nu_h$ are set to $(\Delta x / \Delta z)^2\kappa_h$
and $(\Delta x / \Delta z)^2\nu_h$ respectively.
All simulations are run until equilibrium is
established, and the last 10,000 rotation periods are used for diagnostic
purposes.

The Boussinesq model (Eq.~\ref{eq:boussinesq}) is integrated with
Oceananigans.jl \cite{ramadhan2020oceananigans, silvestri2025gpu}, a
high-performance ocean general circulation model coded in Julia. We use a
staggered Arakawa C-grid \cite{arakawa1977computational} and a third-order
Runge-Kutta method for time integration \cite{le1991improvement}. The advection
terms are calculated with a fifth-order WENO (weighted essentially
non-oscillatory advection) scheme \cite{shu2009high,silvestri2024new}. The 
topography at the top is represented by an immersed boundary \cite{mittal2005immersed}. A
non-hydrostatic solver is used and we use the conjugate gradient method
preconditioned by a three-dimensional Fast Fourier Transform to calculate $P$ in
Eq.~\ref{eq:boussinesq}.

\section{Phenomenology of ocean circulation}
 \label{sec:solutions}

In this section, we describe the ocean circulation and associated transport
characteristics from simulations ``h4+'' and ``h4-''
(see Table~\ref{tab:simulation} for the
parameters of this simulation), which use positive and negative $\Delta b$
, respectively, and both feature the largest topographic
height difference $\Delta H$. This case exemplifies the mechanisms that
govern ocean circulation in all simulations in our study. These mechanisms are
consistent with previous work prescribing lateral buoyancy gradients over flat
boundaries \cite{kang2022does, zhang2024ocean}, although tilted topography can
significantly influence circulation strength. This motivates the scaling
framework developed in Sec.~\ref{sec:scaling}.

\begin{figure}
    \centering
    \includegraphics[width=\linewidth]{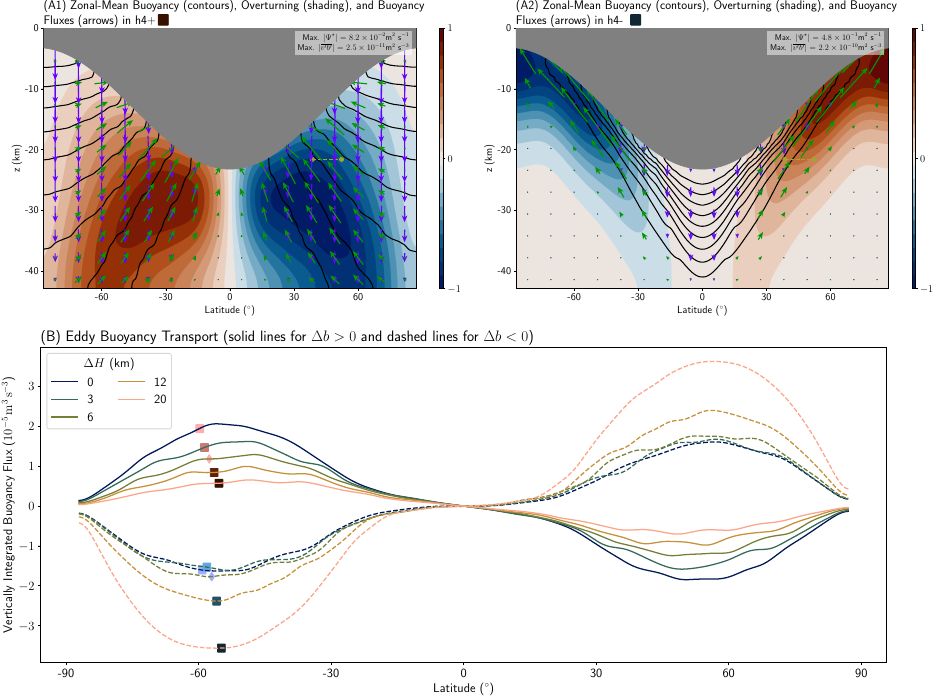}
    \caption{
        Ocean circulation and buoyancy transport in simulations h4$^+$ and h4$^-$ (see
        Table~\ref{tab:simulation} for parameters).
        Panel~A1 corresponds to h4$^+$; Panel~A2 to h4$^-$.
        In Panels A1 and A2, solid contours show buoyancy
        (contour interval = $|\Delta b|/10$),
        with buoyant water overlying denser water. Green and
        purple arrows denote eddy buoyancy flux $(\overline{v'b'},
        \overline{w'b'})$ and diffusive flux $(-\kappa_h \partial_y
        \overline{b}, -\kappa_v \partial_z \overline{b})$, normalized by the
        maximum $\overline{v'b'}$ (listed top right). The equivalent overturning
        circulation, diagnosed from Eq.~\ref{eq:diagnose-psi-star}, is shown by
        $\psi^\star$ shading (normalized by its maximum; values listed).
        Negative $\psi^\star$ (blue) indicates counterclockwise overturning;
        positive (red) indicates clockwise overturning. Panel~B shows vertically
        integrated buoyancy transport: solid lines for h1$^+$--h4$^+$ and c1$^+$ (varying
        $\Delta H$; legend), dashed lines for h1$^-$--h4$^-$ and c1$^-$.
    }
    \label{fig:phenomenology}
\end{figure}

The flow is dominated by baroclinic eddies
(Fig.~\ref{fig:phenomenology}(A1,A2)). Eddies extract available potential energy from
the large-scale meridional buoyancy gradient through baroclinic instability. Zonal
jets also emerge as eddy kinetic energy is arrested at the Rhines scale by the
$\beta$-effect \cite{Rhines-1975:waves, Rhines-1979:geostrophic}, consistent
with prior findings \cite{heimpel2007turbulent, bire2022exploring,
zhang2024ocean}. These eddies generate buoyancy fluctuations that drive a
down-gradient meridional buoyancy flux $v'b'$
which converges towards the equator for $\Delta b > 0$ and towards the poles
for $\Delta b < 0$ (Fig.~\ref{fig:phenomenology}).

The transport characteristics of ocean circulation, which is the focus of this
study, can be described by an eddy-driven overturning. Similarly to
\citeA{kang2022does, zhang2024ocean}, eddies dominate meridional heat transport,
while the Eulerian mean flow contributes negligibly.
Since turbulent flow is
nearly adiabatic, the buoyancy flux (green arrows) aligns with isopycnals
(contours) (Fig.~\ref{fig:phenomenology}(B)):
\begin{linenomath*}
\begin{equation}
    \frac{\overline{w'b'}}{\overline{v'b'}}
    = -\frac{\partial_y \overline{b}}{\partial_z \overline{b}}.
    \label{eq:flux-being-skew}
\end{equation}
\end{linenomath*}
This supports using a
residual-mean streamfunction $\psi^\star$ to describe eddy transport
\cite{plumb2005transformed, zhang2024ocean}:
\begin{linenomath*}
\begin{equation}
    \psi^\star = -\frac{\overline{w'b'}}{\partial_y \overline{b}},
    \label{eq:diagnose-psi-star}
\end{equation}
\end{linenomath*}
which is equivalently expressed as
$\psi^\star = \overline{v'b'} / \partial_z \overline{b}$
(using Eq.~\ref{eq:flux-being-skew}). In practice, we use
Eq.~\ref{eq:diagnose-psi-star} because it remains well-defined throughout the
domain. For $\Delta b > 0$, the overturning sinks at poles, the
source of cold water, resulting in a counter-clockwise overturning (positive
$\psi^\star$) in the Southern Hemisphere; for $\Delta b > 0$, the overturning
sinks at the equator instead, reversing the direction of the overturning.
We will see in Sect.~\ref{sec:scaling:eddy} that the strength of overturning
depends on the buoyancy gradient that energizes eddies and the background
rotation of the moon, and a scaling for $\psi^\star$ will be given.

\begin{figure}
    \centering
    \includegraphics[width=\linewidth]{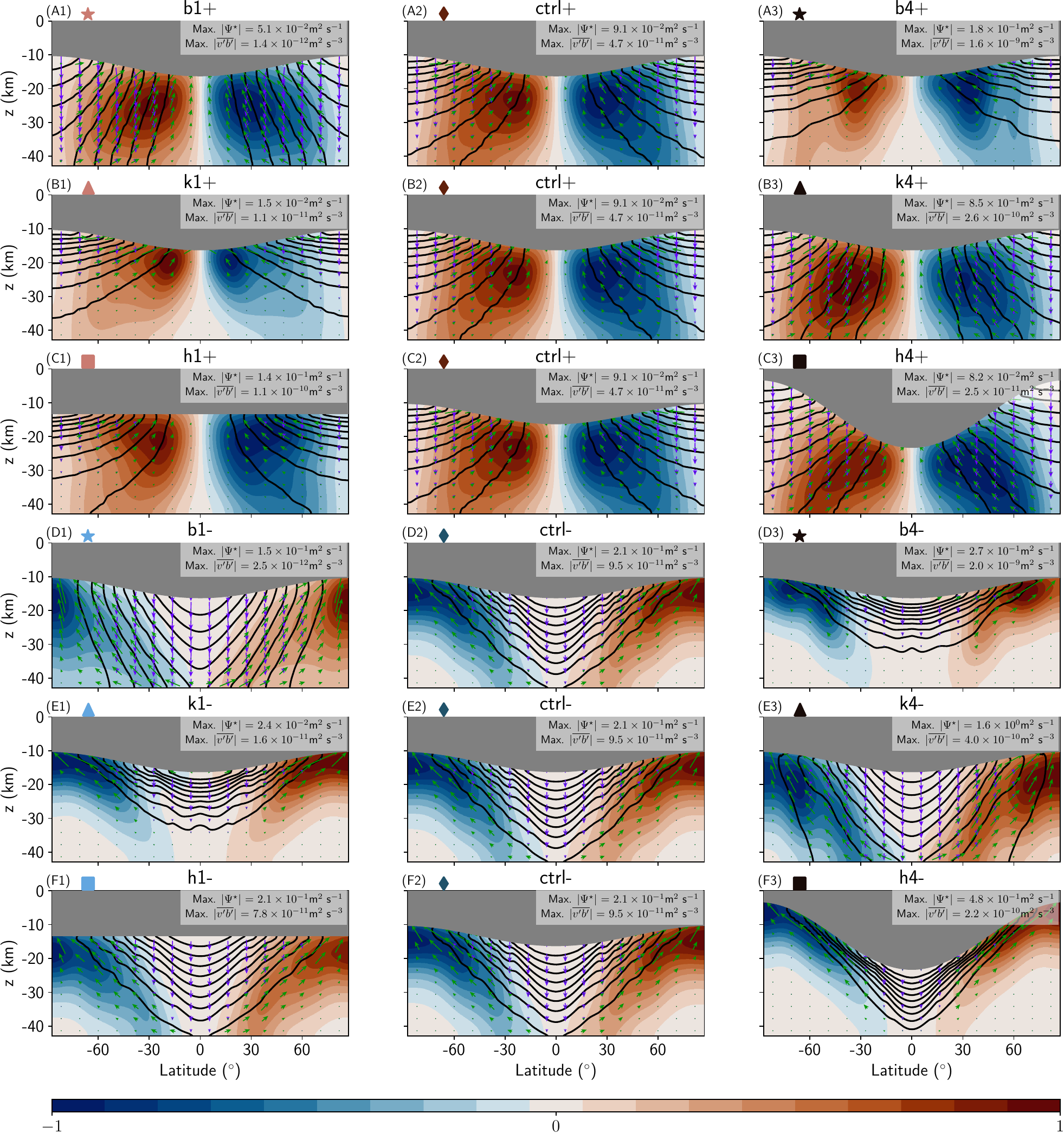}
    \caption{Zonal-mean buoyancy and buoyancy flux patterns from our numerical
    simulations. Each panel corresponds to one simulation (see
    Table~\ref{tab:simulation} for parameters) and shows the same variables
    as Fig.~\ref{fig:phenomenology}(A) The middle column shows the control
    experiment for $\Delta b>0$ (top 3 panels) and $\Delta b<0$ (bottom 3
    panels). For convenience the panels presenting the control are repeated to
    ease comparison as parameters are changed moving left and right.
    }
    \label{fig:zonal-mean}
\end{figure}

Vertical diffusion is essential to maintain the circulation. Unlike the
Earth's ocean, where wind-driven Ekman pumping transfers surface buoyancy gradients
downward, icy moon oceans lack such forcing. Instead, vertical diffusion must
transport the buoyancy gradient from the ice-ocean interface to the interior.
Energetically, a positive (upward) $\overline{w'b'}$ is required to maintain the
eddy field, which must be balanced by a downward diffusive buoyancy
flux--precisely what we observe in simulations
(Fig.~\ref{fig:phenomenology}(A) and Fig.~\ref{fig:zonal-mean}).

For positive $\Delta b$ (as in the case of ``h4+''),
the sloped topography creates a thicker stratified layer
below the poles (Fig.~\ref{fig:phenomenology}(A1)). In this case,
buoyant water sits higher at the poles than denser water at the equator,
weakening circulation: vertical diffusion must penetrate deeper to transport
buoyancy into the interior and trigger baroclinic instability. The decreasing
meridional buoyancy transport with increasing $\Delta H$
(Fig.~\ref{fig:phenomenology}(B)) supports this interpretation.
This idea becomes further developed in Sect.~\ref{sec:scaling:heat}
where a formula for the penetration depth is presented which depends on the strength of vertical diffusion (represented
by $\kappa_v$) and the eddy overturning streamfunction (represented
 by the maximum of $|\psi^\star|$), following \citeA{kang2022does,
zhang2024ocean}

For negative $\Delta b$ (as in the case of ``h4-''), the sloped
topography provides a new pathway to energize the eddies. In this case, buoyant
water at the equator can move upward along the tiled ocean top toward the poles.
The upward eddy buoyancy $\overline{w'b'}$, which must be balanced by downward vertical diffusion in the previous case, can now be balanced
by an exchange of buoyancy with the tilted upper boundary layer.
As a result, the meridional buoyancy flux strengthens with
increasing $\Delta H$ (Fig.~\ref{fig:phenomenology}(B)).
In addition, the depth of the stratified layer shrinks, leaving the major part of the ocean cold and almost free of baroclinic eddies.
Comparing our simulation ``h4+'' ($\Delta b > 0$) and ``h4-'' ($\Delta b < 0$),
the former has stratified layer (isopycnals shown by contours) and overturning circulation (meridional streamfunction shown by shading
Fig.~\ref{fig:phenomenology}(A1)) extending from the ocean top almost to the
bottom, but in the latter, both the stratified layer and overturning circulation
are shallow (Fig.~\ref{fig:phenomenology}(A2)).

Fig.~\ref{fig:zonal-mean} shows how the equilibrium solutions change as the
mixing (rows A and D), buoyancy parameters (rows B and E), and topography (rows
C and F) are varied relative to the control. Their effects on the equilibrium
flow are discussed in Sect.~\ref{sec:results}, after establishing the scaling
relations in Sect.~\ref{sec:scaling}.

\section{Scaling Theory}
    \label{sec:scaling}

In this section, we derive the scaling laws that govern how 1) the meridional
buoyancy (heat) flux by baroclinic eddies $F_h$ and 2) the equilibrium buoyancy
(temperature) distribution vary with planetary parameters and ocean diffusivity.
To achieve this goal, we need to take two steps. First (in
Sect.~\ref{sec:scaling:heat}), we determine the equilibrium buoyancy distribution,
characterized by the isopycnal slope $s$, since downward buoyancy diffusion
balances upward buoyancy transport by eddies. Second (in
Sect.~\ref{sec:scaling:eddy}), we determine the transport efficiency of
baroclinic eddies, characterized by the residual overturning streamfunction
$\psi^\star$, for a given isopycnal distribution $s$.

 \begin{figure*}
    \centering
    \includegraphics[width=\linewidth]{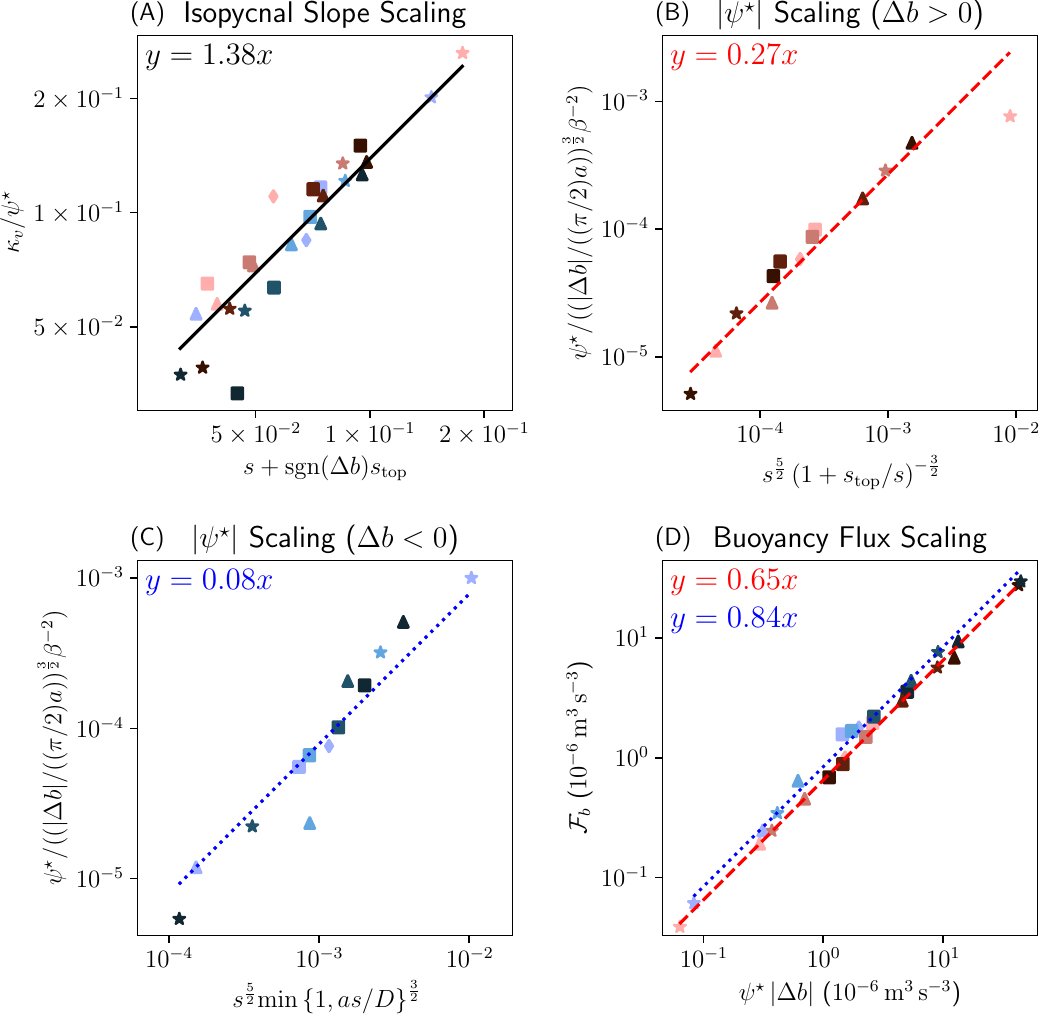}
    \caption{Tests of scaling theory against numerical simulations. Panel~(A)
        shows a scaling for the isopycnal slope (Eq.~\ref{eq:scaling-kappa-over-psi}).
        Panel~(B) and Panel~(C) show the scaling
        for the eddy-driven overturning $\psi^\star$ for positive
        and negative $\Delta b$, respectively
        (Eq.~\ref{eq:scaling-psi}).
        Panel~(D) shows the scaling for the vertical integrated meridional
        buoyancy flux (proportional to heat flux).
        Here, $s$ represents the slope of the median isopycnal, $s_m$;
        $s_\mathrm{top}\equiv \Delta H / ((\pi/2)a)$ is the slope of the top
        topography; $\psi^\star$ represents the mean overturning diagnosed
        using Eq.~\ref{eq:diagnose-psi-star};
        $\mathrm{sgn}(\Delta b)$ is the sign of the prescribed
        equator-to-pole buoyancy difference $\Delta b$;
        $\mathcal{F}_b$ is the maximum of the vertical integrated meridional
        buoyancy flux. In Panel~(B) and (C),
        $\psi^\star$ is normalized using the buoyancy difference
        $\Delta b$, the moon radius $a$,
        and the $\beta$ factor ($\beta \equiv 2\Omega / a$). The
        line and the formula at the upper left corner of each panel shows
        the best fit; the red dashed line and blue dotted
        lines represent the fit for positive and negative $\Delta$
        cases, respectively.
        The marker and color of each simulation is the same as 
        Fig.~\ref{fig:zonal-mean}. The parameters used in these simulations are
        set out in Table~\ref{tab:simulation}.
    }
    \label{fig:motivate-scaling}
\end{figure*}
 
\subsection{Buoyancy distribution -- the competition between diffusion and
    baroclinic eddies}
    \label{sec:scaling:heat}

Here, we try to determine the equilibrium buoyancy distribution assuming that we
know the circulation strength. For simplicity, this buoyancy distribution is
characterized by one parameter, the isopycnal slope
$s\equiv |\partial_y \overline{b}/\partial_z \overline{b}|$. Following
\citeA{lobo2021pole, kang2022icy, zhang2024ocean}, $s$ in the equilibrium state
is determined by two competing effects. On the one hand, vertical diffusion,
controlled by vertical diffusivity $\kappa_v$, brings the buoyancy gradient into
the interior of the ocean, steepening the isopycnal and increasing the available
potential energy of the ocean \cite{jansen2022energetics}. On the other hand,
baroclinic eddies tap into this energy source and tend to flatten out the
isopycnals \cite{marshall2003residual,ito2008control}. 
Since the turbulent mixing process is close to adiabatic, the eddy-induced heat
transport should be directed along the isopycnals, and so the transport can be
equivalently represented by an overturning circulation $\psi^\star$:
\begin{linenomath*}\begin{align}
    \overline{v'b'} &\sim -\mathrm{sgn}(\Delta b) |\psi^\star| \partial_z \overline{b}
    \label{eq:scaling-vb-psi}
    \\
    \overline{w'b'} &\sim |\psi^\star| |\partial_y \overline{b}|
    \label{eq:scaling-wb-psi}
\end{align}\end{linenomath*}
where $\mathrm{sgn}(\Delta b)$ represents the sign of the equator-pole
buoyancy gradient. A positive $\Delta b$ corresponds to having denser water near
the equator and vise versa. Note that the sign of $\overline{v'b'}$ and all
variables in Sect.~\ref{sec:scaling:heat} is given for the Northern Hemisphere.

The total heat flux in the direction perpendicular to the top topography 
must vanish at equilibrium. In other words,
$(\overline{v'b'}, \overline{w'b'})$ is parallel to the top topography:
\begin{linenomath*}\begin{equation}
    \underbrace{|\psi^\star| |\partial_y \overline{b}|}_{\overline{w'b'}}
    - \kappa_v\partial_z b
    \sim s_\mathrm{top}
    \underbrace{(-\mathrm{sgn}(\Delta b)|\psi^\star| \partial_z \overline{b})}_{
        \overline{v'b'}
    }
    \label{eq:vertical-heat-balance}
\end{equation}\end{linenomath*}
where $s_\mathrm{top}$ is the slope of the top topography. Dividing
both sides of Eq.~\ref{eq:vertical-heat-balance} by $\overline{w'b'}$
and using $s = |\partial_y \overline{b}| / \partial_z \overline{b}$,
we get a formula for the isopycnal slope $s$:
\begin{linenomath*}\begin{equation}
    s + s_\mathrm{top}\mathrm{sgn}(\Delta b) \sim \kappa_v / |\psi^\star|
    \label{eq:scaling-kappa-over-psi}
\end{equation}\end{linenomath*}

This tells us that when diffusion dominates ($\kappa_v>\psi^\star$), the isopycnal slope
$s$ would steepen and vise versa, consistent with the physical intuition presented at the beginning of this section. Furthermore, the topography slope
$s_\mathrm{top}$ decreases $s$ when they tilt in the opposite direction when
$\Delta b>0$ and increases $s$ when $\Delta b<0$.
\ref{appendix:isopycnal-slope-scaling} presents a rigorous derivation of
Eq.~\ref{eq:scaling-kappa-over-psi} using the integral form of the heat flux
balance.

We test Eq.~\ref{eq:scaling-kappa-over-psi} against our simulations 
(Fig.~\ref{fig:motivate-scaling}(A)). To
measure $s$, we measure the slope of the median isopycnal $s_\mathrm{m}$, by
tracking its latitudinal and depth extent.
To measure the amplitude of the eddy
driven-overturning $\psi^\star_m$, we project the vertical heat flux
$\overline{w'b'}$ onto the meridional buoyancy gradient $\partial_y
\overline{b}$:
\begin{linenomath*}\begin{equation}
    \psi^\star_\mathrm{m} = \frac{
        \int_\mathcal{D} |\overline{w'b'}\partial_y\overline{b}|\mathrm{d}y\mathrm{d}z
    }{
        \int_\mathcal{D} \left(\partial_y\overline{b}\right)^2\mathrm{d}y\mathrm{d}z
    }
    \label{eq:psi-measured}
\end{equation}\end{linenomath*}
where $\mathcal{D}$ represents the ocean domain of $(y, z)$. This form is similar
to how we diagnose the spatial distribution of $\psi^\star$ (Eq.~\ref{eq:diagnose-psi-star}).
Here, we use an average over the domain to
get the overall overturning strength. 
The topographic slope $s_\mathrm{top} \equiv \Delta H
/ ((\pi / 2)a)$. In Fig.~\ref{fig:motivate-scaling}(A), we show that the
right-hand side and the left-hand
side of Eq.~\ref{eq:scaling-kappa-over-psi} scale with each other, suggesting
that this formula captures how diffusion and baroclinic eddies jointly affect the
slope of isopycnals. The fitted ratio $c_s = 1.38$ of the proportional
relationship (Fig.~\ref{fig:motivate-scaling}(A)) may be a result
of our choice of how to diagnose $\psi^\star_m$ and $s_m$. We include
the factor of $c_s$ in our scaling framework for consistency.

\subsection{Efficiency of baroclinic eddy transport $\psi^\star$}
    \label{sec:scaling:eddy}

In this section, we present a scaling law for the eddy-driven overturning
$\psi^\star$ given an isopycnal slope. 

Following the Gent-McWilliams closure \cite{Gent-Mcwilliams-1990:isopycnal}, the overturning streamfunction can be written as the product of the eddy diffusivity $\kappa_e$ and the isopycnal slope $s$,
\begin{linenomath*}\begin{equation}
    |\psi^\star| = \kappa_e s.
    \label{eq:psi-scales-with-ke}
\end{equation}\end{linenomath*}
We can verify
$|\overline{v'b'}| \sim \kappa_e |\partial_y \overline{b}|$ from
Eq.~\ref{eq:scaling-vb-psi} as a way to motivate
Eq.~\ref{eq:psi-scales-with-ke}.

To obtain scaling laws for $\psi^\dagger$, we first need to know how $\kappa_e$
depends on the buoyancy profile $b$. Following the mixing-length theory,
$\kappa_e$ can be expressed as the product of the ' characteristic length
scale of eddies $L_e$ and their characteristic velocity $V_e$.
\begin{linenomath*}\begin{equation}
  \kappa_e=kL_eV_e
  \label{eq:mixing-length}
\end{equation}\end{linenomath*}
where $k\sim 0.25$ is a constant. 

To estimate the eddy length scale $L_e$ and the velocity scale $V_e$, we follow the
pioneering work of \citeA{Held-Larichev-1996:scaling} and
\citeA{Gallet-Ferrari-2021:quantitative}. The physical picture presented in these works is as follows. On a $\beta$-plane, baroclinic eddies, once
generated, cascade to increasingly larger scales until reaching the zonal 
scale $L_\beta$. Beyond $\L_\beta$, most of the kinetic energy of the eddy is in the form of
zonal jets, which do not transport buoyancy (heat) meridionally. Therefore, the 
characteristic size of the eddy $L_e$ should scale with $L_\beta$:
\begin{linenomath*}\begin{equation}
  L_e\sim L_\beta=\sqrt{V_e/\beta}.
  \label{eq:Le-Lbeta}
\end{equation}\end{linenomath*}

At equilibrium, the rate at which baroclinic energy is generated
$\left.\dot{E}\right._{\rm bci}$ should be equal to
$\left.\dot{E}\right._{L_\beta}$, the rate of energy transfer across $L_\beta$.
The latter can be expressed as a function of the characteristic velocity $V_e$
and characteristic length $L_\beta$
\cite{Galperin-Sukoriansky-Dikovskaya-2008:zonostrophic}:
\begin{linenomath*}\begin{equation}
  \label{eq:cascade-rate}
  \left.\dot{E}\right._{L_\beta}=V_e^3/L_\beta , 
\end{equation}\end{linenomath*}
By equating $\left.\dot{E}\right._{\rm bci}$ and
$\left.\dot{E}\right._{L_\beta}$, \citeA{Held-Larichev-1996:scaling} arrived at scaling laws for $L_e$ and $V_e$.

To evaluate $\left.\dot{E}\right._{\rm bci}$, \citeA{Held-Larichev-1996:scaling}
and \citeA{Gallet-Ferrari-2021:quantitative} a two-layer
quasi-geostrophic (QG) system is considered. Although similar analyzes
are possible for a tilted boundary \cite{Mechoso-1980:baroclinic}, here we
consider the energy release by baroclinic eddies in a 3D system:
\begin{linenomath*}\begin{equation}
  \label{eq:energy-generation}
  \left.\dot{E}\right._{\rm bci}=\overline{w'b'}=s|\overline{v'b'}|=s V_eL_e \overline{|\partial_yb|},
\end{equation}\end{linenomath*}
where $\overline{\partial_yb}$ denotes the meridional buoyancy gradient averaged
from the surface, where the buoyancy gradient (and hence available potential energy,
APE) is strongest, to the level where baroclinic instability actually occurs
.

Setting Eq.~\eqref{eq:cascade-rate} and Eq.~\eqref{eq:energy-generation} equal to each other yields
\begin{linenomath*}\begin{equation}
  \label{eq:scaling-V-Lbeta}
  V_e=s\overline{|\partial_yb|}\beta^{-1},\ L_e\sim L_\beta=(s\overline{|\partial_yb|})^{1/2}\beta^{-1},
\end{equation}\end{linenomath*}
which, when combined with Eq.~\ref{eq:mixing-length}, leads to
\begin{linenomath*}\begin{equation}
    \kappa_e \sim \left(s\overline{|\partial_y b|}\right)^\frac{3}{2}\beta^{-2}.
    \label{eq:kappa-e-scaling}
\end{equation}\end{linenomath*}
This scaling shows how the meridional buoyancy gradient and the isopycnal
slope affect the eddy transport efficiency.

To calculate $|\partial_y b|$, the meridional buoyancy gradient must be averaged
over the region where baroclinic instability occurs and develops, which
corresponds roughly to the vertical extent of the fastest-growing mode.
When $\Delta b>0$, polar fluid is denser than equatorial fluid, the
potential vorticity (PV) gradient near the upper surface is of the opposite sign
to that in the interior (dominated by the planetary $\beta$), satisfying the Charney-Stern
criterion for instability \cite{charney1962stability}. Since the upper surface
is also where APE is
concentrated, APE release does not depend on conditions at the bottom of the ocean, where buoyancy is constant. This suggests that $\overline{\partial_yb}$
should be averaged over the upper layer, where surface buoyancy anomalies penetrate, giving the result that
\begin{linenomath*}\begin{equation}
  \label{eq:by-positive}
  \overline{|\partial_yb|} \sim \frac{|\Delta b|}{(\pi / 2)a} \frac{1}{1+s_{\rm top}/s}
  \quad \mathrm{for}\quad\Delta b>0.
\end{equation}\end{linenomath*}

In contrast, when $\Delta b<0$, the PV gradient near the bottom boundary surface
is of opposite sign to the interior, so the Charney-Stern instability criterion
is only satisfied in the lower part of the ocean. Since there is very little APE
in the bottom ocean, only modes that extend the entire ocean depth can tap into
the APE source. This justifies averaging $\partial_y b$ across the entire ocean
\begin{linenomath*}\begin{equation}
  \label{eq:by-negative}
  \overline{|\partial_yb|}
  \sim \frac{|\Delta b|}{(\pi / 2)a}\frac{1}{D}\mathrm{min}\left\{D,
      as\right\} \quad \mathrm{for}\quad \Delta b<0.
\end{equation}\end{linenomath*}

Substituting Equation~\ref{eq:by-positive} or \ref{eq:by-negative}
into Equations~\ref{eq:kappa-e-scaling} and \ref{eq:psi-scales-with-ke}, we get
\begin{linenomath*}\begin{equation}
    |\psi^{\star}| = \left\{\begin{tabular}{ll}
        $c_\psi^{+}\left(\frac{|\Delta b|}{(\pi / 2)a}\right)^\frac{3}{2}
        \beta^{-2}
        s^\frac{5}{2}
        \left(1 + s_\mathrm{top}/s\right)^{-\frac{3}{2}}$
        & for $\Delta b > 0$;
    \\
        $c_\psi^{-}\left(\frac{|\Delta b|}{(\pi / 2)a}\right)^\frac{3}{2}
        \beta^{-2}
        s^\frac{5}{2}
        \mathrm{min}\left\{1, as/D \right\}^\frac{3}{2}$
        & for $\Delta b < 0$,
    \\
\end{tabular} \right.
\label{eq:scaling-psi}
\end{equation}\end{linenomath*}
where $c_\psi^{+}$ and $c_\psi^{-}$ are constants that can be obtained
by regression analysis.

We test Eq.~\ref{eq:scaling-psi} against our simulations
(Fig.~\ref{fig:motivate-scaling}(B,C)). The measured isopycnal slope and
 overturning stream function use the same definitions as
Sect.~\ref{sec:scaling:heat}. Although $\beta = (2\Omega / a) \sin{\theta}$
varies with latitude $\theta$, for simplicity, we set $\beta$ at $2\Omega / a$
when performing the scaling analysis, absorbing the influence of latitudinal dependence into $c_\psi^+$ or $c_\psi^-$ in Equation~\ref{eq:scaling-psi}. Panels~(B) and (C) show a good agreement between simulations and our scaling, yielding $c_\psi^+ = 0.27$ and $c_\psi^{-} = 0.08$.

\subsection{Prediction of ocean heat transport}

Combining Equations~\ref{eq:scaling-kappa-over-psi} and \ref{eq:scaling-psi},
one can solve for the isopycnal slope $s$. The combined equation
is:
\begin{linenomath*}\begin{align}
    s^{4}\left(s + s_\mathrm{top}\right)^{-\frac{1}{2}} &=
        \left(c_s c_\psi^{+}\right)^{-1}
        \kappa_v
        \left(\frac{|\Delta b|}{(\pi / 2)a}\right)^{-\frac{3}{2}}
        \beta^{2} & \mathrm{for}\quad \Delta b  > 0;
    \label{eq:scaling-s-positive}
    \\
    \left(s - s_\mathrm{top}\right)s^\frac{5}{2}
    \mathrm{min}\left\{1, as/D\right\}^\frac{3}{2} &=
        \left(c_s c_\psi^{-}\right)^{-1}
        \kappa_v
        \left(\frac{|\Delta b|}{(\pi / 2)a}\right)^{-\frac{3}{2}}
        \beta^{2} & \mathrm{for}\quad \Delta b  < 0.
    \label{eq:scaling-s-negative}
\end{align}\end{linenomath*}
One can verify that the left hand side monotonously increases with $s$.
For $\Delta b < 0$, $s$ must be greater than $s_\mathrm{top}$ for the left
hand side to be valid.

Once $|\psi^\star|$ is known, the implied meridional buoyancy transport
$\mathcal{F}_b$ may be estimated by the following:
\begin{linenomath*}\begin{equation}
\mathcal{F}_b = c_F|\psi^\star| |\Delta b|.
\label{eq:scaling-f}
\end{equation}\end{linenomath*}
To measure $\mathcal{F}_b$ in a particular simulation, we first calculate
the vertically integrated buoyancy flux as a function of $y$
and then take the maximum as the characteristic value
that we present in Fig.~\ref{fig:motivate-scaling}(D) and
Fig.~\ref{fig:heat-flux}.
Here, $c_F$ can be different for positive and negative $\Delta b$ because the
sign of $\Delta b$ affects the direction in which the isopycnals tilt and therefore determines whether the stratification is centered at low or high latitudes.
We fit these two cases separately and find good agreement between
$\mathcal{F}_b$ and $|\psi^\star_m| \Delta b$ (Fig.~\ref{fig:motivate-scaling}(D)).
The fitted parameters are $c_F^+ = 0.65$ and $c_F^- = 0.84$ for positive and
negative $\Delta b$, respectively.

\section{Effects of key parameters}
    \label{sec:results}

\begin{figure*}
    \centering
    \includegraphics[width=\linewidth]{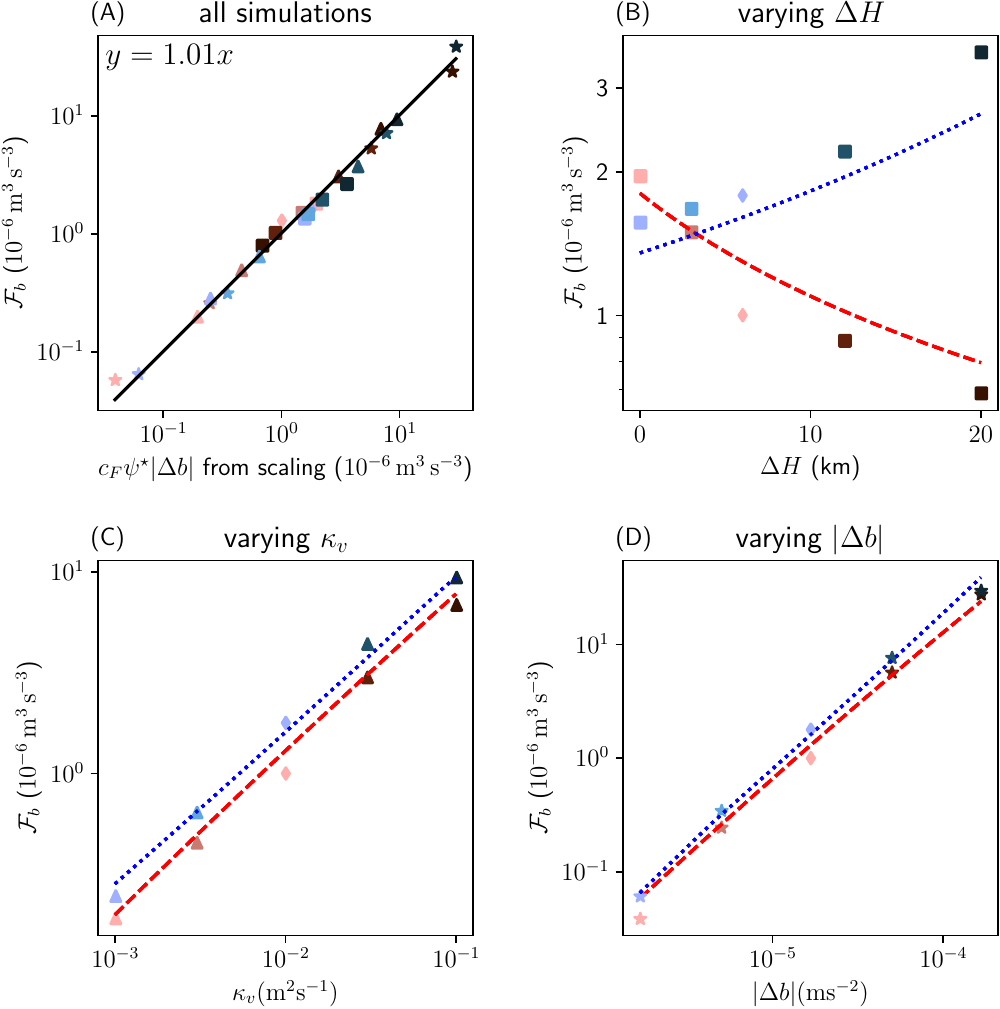}
    \caption{Meridional buoyancy fluxes in our simulations
        compared against predictions from our scaling framework
        (Equations~\ref{eq:scaling-s-positive}, \ref{eq:scaling-s-negative},
        \ref{eq:scaling-psi}, and \ref{eq:scaling-f}).
        Panel~(A) plots the predicted vertically integrated meridional buoyancy
        transport $\mathcal{F}_b$ against the measured value (defined
        as the maximum along all latitudes) from each of our simulations. The solid line and the formula on the left upper corner show the best fit,
        and is almost the 1:1 line. Panels~(B,C,D) show how $\Delta H$,
        $\kappa_v$, and $|\Delta b|$ influences $\mathcal{F}_b$: the dashed
        and dotted lines represent the predictions of our scalings
        for positive and negative $\Delta b$, respectively. The parameters used
        in these simulations are tabulated in Table~\ref{tab:simulation}.
    }
    \label{fig:heat-flux}
\end{figure*}

We test our scaling for the vertically integrated meridional buoyancy flux,
$\mathcal{F}_b$, using 26 numerical simulations spanning a range of $\Delta H$,
$\kappa_v$, and $\Delta b$ (Fig.~\ref{fig:heat-flux}). All data points fall
along the 1:1 line, indicating general good agreement between simulations 
and scaling prediction (Fig.~\ref{fig:heat-flux}(A)).

To understand how each individual parameter ($\Delta H$, $\kappa_v$, and $\Delta
b$) influences the buoyancy transport $\mathcal{F}_b$ and the underlying
physical mechanisms, we revisit the conceptual framework. In our model setup
(Fig.\ref{fig:model-setup}), buoyancy varies along the upper surface, but unless
this variation induces a non-zero meridional buoyancy gradient, there is no
available potential energy to drive a circulation. When the water-ice interface
is flat, as assumed in previous studies \cite{kang2022icy, zhang2024ocean},
vertical diffusion is essential to initiate baroclinic instability and
allow heat transport. In this regime, the meridional buoyancy flux
$\mathcal{F}_b$ scales as $\kappa_v^{5/7}$. This prediction can be
derived by solving Equations~\ref{eq:scaling-s-positive}, \ref{eq:scaling-psi},
and \ref{eq:scaling-f}, and the resulting $5/7$ power law is confirmed by our
numerical simulations (Fig.~\ref{fig:heat-flux}(C)).

When topography is introduced, the above physical picture may be subject to
changes. If denser fluid lies beneath thicker ice ($\Delta b > 0$), the
stratification (measured by $\partial_z \overline{b}$) beneath thin ice is
weakened due to the increased distance from the water-ice interface to a
specific isopycnal. This reduced stratification
weakens the vertical diffusive flux and thereby the meridional buoyancy
transport, since the vertical and meridional buoyancy fluxes must balance in
equilibrium (Eq.~\ref{eq:vertical-heat-balance}). This trend is evident in our
numerical simulations, as shown in Fig.~\ref{fig:zonal-mean}(C) and
Fig.~\ref{fig:heat-flux}(B).

In contrast, when the denser fluid lies beneath thinner ice ($\Delta b < 0$), the
isopycnals bend toward the equator and are compressed by the protruding ice,
enhancing vertical diffusion and meridional heat transport
(Fig.~\ref{fig:zonal-mean}(F)). In this configuration, unlike the case with
$\Delta b > 0$, the dynamics is not solely energized by downward diffusion.
Instead, surface buoyancy forcing injects available potential energy into the
system by densifying the ocean from above \cite{jansen2022energetics}. As a
result, even in the limit $\kappa_v \rightarrow 0$, the meridional buoyancy
transport does not vanish. This behavior is captured in
Eq.~\ref{eq:scaling-psi}: substituting $s = s_\mathrm{top} > 0$ yields a
non-zero $\psi^\star$. The dependence of $\mathcal{F}_b$ on $\Delta H$ in the
case of negative $\Delta b$ is shown in Fig.\ref{fig:heat-flux}(B) (blue line),
which is in good agreement with the simulation results (blue dots).

Increasing the equator-to-pole buoyancy contrast $|\Delta b|$ initially
strengthens the overturning circulation, lifting isopycnals upward
(Fig.~\ref{fig:zonal-mean})(A,D)). According to
the vertical buoyancy flux balance (Eq.~\ref{eq:vertical-heat-balance}), this
intensifies the vertical diffusive flux and, consequently, the meridional heat
transport. After some mathematical manipulation, one can show that
$\mathcal{F}_b \sim |\Delta b|^{10/7}$, consistent with the scaling laws derived
in \citeA{kang2022icy, zhang2024ocean}. This dependence is also consistent with our simulations (Fig.~\ref{fig:heat-flux}(C)).

$\Delta b$ is a prescribed constant in our simulations, but on realistic icy
moons it is correlated with many other parameters, including $\Delta H$,
$\alpha$, and $g$ (which increases with the moon radius $a$). We apply
our scaling to realistic icy moons, including Enceladus, Europa, and Titan,
to predict the ocean heat transport there in Sect.~\ref{sec:application}.

\section{Implication for Icy Moons}

\label{sec:implications}

\subsection{Role of topography on realistic icy moons}
\label{sec:role-of-topography}

\begin{figure*}
    \centering
    \includegraphics[width=\linewidth]{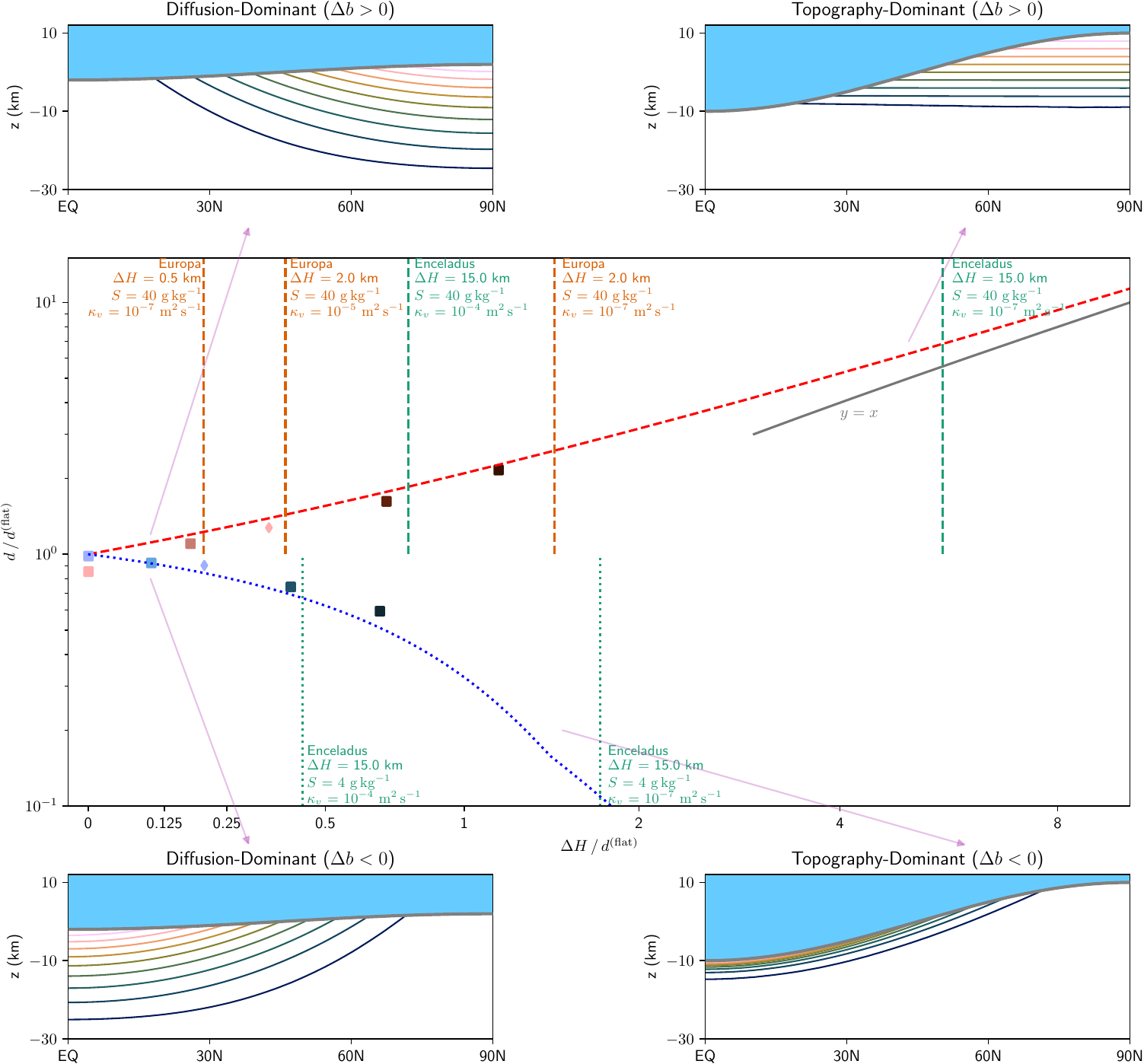}
    \caption{Effect of top topography on ocean stratification. Here,
        $d^\mathrm{(flat)}$ is the penetration depth assuming a flat ocean top,
        computed using the scaling of \citeA{kang2022different,zhang2024ocean}
        (which is consistent with this study when $s_\mathrm{top}=0$); $d$ is the actual
        stratified layer thickness (see Sect.~\ref{sec:role-of-topography} for
        definitions). Red dashed and blue dotted lines show $d/d^\mathrm{(flat)}$
        as a function of $\Delta H/d^\mathrm{(flat)}$ for $\Delta b > 0$ and
        $\Delta b < 0$, respectively. When $\Delta H \to 0$, $d \approx
        d^\mathrm{(flat)}$, stratification is set by vertical diffusion
        (upper left and lower left sketches). As $\Delta H$ increases, $d$ grows
        for $\Delta b>0$ but shrinks for $\Delta b<0$. In the large $\Delta
        H/d^\mathrm{(flat)}$ limit and when $\Delta b>0$, the stratified layer exists between the minimum and maximum upper topographic depths and the isopycnals are nearly flat (upper right). In contrast when $\Delta b<0$, the stratified layer becomes thin and hugs the topography
        (lower right). Vertical lines indicate $\Delta H/d^\mathrm{(flat)}$ typical of Enceladus and Europa, for given $\Delta H$, ocean salinity $S$, and
        vertical diffusivity $\kappa_v$; their position (upper or lower)
        reflects the sign of $\Delta b$ set by $S$ which controls the sign of the thermal
        expansion coefficient.
    }
    \label{fig:difference}
\end{figure*}

A key question is to what extent topography influences ocean circulation on
realistic icy moons. To investigate this, we focus on the thickness of the stratified layer, $d$, defined as the maximum vertical distance between the median isopycnal and the ocean surface. This thickness can be expressed as $(\pi/2)a(s_m +
s_\mathrm{top}\mathrm{sgn}(\Delta b))$, where $s_m$ is the slope of the median
isopycnal (introduced in Sect.~\ref{sec:scaling} and shown in
Fig.~\ref{fig:motivate-scaling}) and $s_\mathrm{top}$ is the slope at the top of the ocean.

We examine how the stratified layer thickness $d$ depends on $\Delta H$, the
topographic height difference, and on $d^\mathrm{(flat)}$, the stratified layer
thickness for a flat ocean top. The latter has been extensively studied by
\citeA{kang2022does} and \citeA{zhang2024ocean}. Figure~\ref{fig:difference}
shows $d/d^\mathrm{(flat)}$ plotted against $\Delta H/d^\mathrm{(flat)}$, with
results from simulations (squares) and from our scaling framework (dashed and
dotted lines). In the limit $\Delta H \ll d^\mathrm{(flat)}$, ocean
stratification is set primarily by vertical diffusion and $d$ approaches
$d^\mathrm{(flat)}$. The stratification distribution in this regime is
sketched in the upper-left and lower-left sub-panels for positive $\Delta b$ and
negative $\Delta b$, respectively.

The ratio $\Delta H/d^\mathrm{(flat)}$ indicates how strongly topography affects
ocean stratification and circulation. As $\Delta H$ increases, $d$ increases for
positive $\Delta b$ and eventually approaches $\Delta H$ (red dashed line in
Fig.~\ref{fig:difference}). In this limit, the stratified layer sits just
beneath the topography, with nearly flat isopycnals. In contrast, for negative
$\Delta b$, $d$ decreases as $\Delta H$ increases (blue dotted line in
Fig.~\ref{fig:difference}). For very large $\Delta H$ (lower-right sub-panel of
Fig.~\ref{fig:difference}), the stratified layer collapses into a thin layer
that follows the sloping topography. Although total meridional buoyancy and
heat transport are enhanced, they occur within this thin layer, leaving the deep
ocean largely isolated from circulation.

For Enceladus and Europa, we calculate $\Delta H / d^\mathrm{(flat)}$ under
different assumptions. For Enceladus, with $\kappa_v =
10^{-7}\,\mathrm{m}^{2}\,\mathrm{s}^{-1}$ (comparable to molecular diffusivity;
Table~\ref{tab:icy-moons}, \citeA{soderlund2019ocean}), $d^\mathrm{(flat)}$ is
so small that $\Delta H$ (assumed to be 15 km) exceeds it by several times for
both positive and negative $\Delta b$. In this case, the ocean resembles the
``topography-dominant'' regime (upper and lower right sub-panels). Increasing
$\kappa_v$ to $10^{-4}\,\mathrm{m}^2\,\mathrm{s}^{-1}$ (a typical value for
the Earth's ocean) gives $d^\mathrm{(flat)}$ the same order as $\Delta H$, so both
topography and vertical diffusion become important in establishing stratification and
circulation.

For Europa, $\Delta H/d^\mathrm{(flat)}$ is expected to be smaller. Europa's
$\Delta H$ is at least an order of magnitude lower than Enceladus'
\cite{nimmo2007global}. Assuming $\Delta H = 2$ km and $\kappa_v =
10^{-7}\,\mathrm{m}^2\,\mathrm{s}^{-1}$, we obtain $\Delta H/d^\mathrm{(flat)}
\approx 2$, still in the ``topography-dominant'' regime. However, taking a
smaller $\Delta H$ (e.g., 0.5 km) or a larger $\kappa_v$ greatly reduces $\Delta
H/d^\mathrm{(flat)}$, making topography much less important.

\subsection{Topographic control on the upper limit of heat transport}
    \label{sec:low-diffusivity}

We derive an upper bound on heat transport for the case of positive $\Delta b$.
Using this bound, we examine the limit $\kappa_v \rightarrow 0$, showing that
the meridional heat flux vanishes for positive $\Delta b$ but approaches a
finite value for negative $\Delta b$. This fundamental difference arises from
the underlying energetics.

For positive $\Delta b$, we have the following relationship from
Equation~\ref{eq:scaling-kappa-over-psi}:
\begin{linenomath*}\begin{equation}
    |\psi^\star| \le c_s^{-1} \kappa_v / s_\mathrm{top}
    \label{eq:psi-upper-bound}
\end{equation}\end{linenomath*}
given that $s \ge 0$.

This leads to an upper bound for the meridional heat flux following
Equation~\ref{eq:scaling-f}:
\begin{linenomath*}\begin{equation}
    \mathcal{F}_b \le c_s^{-1}c_F^+ \kappa_v \Delta b / s_\mathrm{top}
    \label{eq:f-b-bound}
\end{equation}\end{linenomath*}

A rigorous derivation from the Boussinesq model (Eq.~\ref{eq:boussinesq})
is given in \ref{appendix:upper-bound}. This upper bound is based on the
fundamental buoyancy flux balance (Sect.~\ref{sec:scaling:heat}) and is
independent of the dynamics of eddies (Sect.~\ref{sec:scaling:eddy}).

The above inequality can be converted into an upper bound of the OHT,
(ocean heat transport in units of $\mathrm{W}$) given our linear
equation (Eq.~\ref{eq:eos}):
\begin{linenomath*}\begin{equation}
    \mathrm{OHT} \le
        (c_s)^{-1} c_F^+
        (\pi^2/2) a^2 g \kappa_v \rho_w^2 c_p
        \frac{\mathrm{d}T_\mathrm{freezing}}{\mathrm{d}p},
    \label{eq:OHT-bound}
\end{equation}\end{linenomath*}
where $\rho_w$ and $c_p$ are the density and heat capacity per mass of seawater
; we multiply $\mathcal{F}_b$, the zonal-mean transport by $\pi a$, the
average circumference of circles of latitudes, to get the total transport by the
three-dimensional flow. Using $\Delta T = \rho_w g \Delta H
({\mathrm{d}T_\mathrm{freezing}}/{\mathrm{d}p})$ and
$s_\mathrm{top} = \Delta H/((\pi/2)a)$, we find that
$\Delta T / s_\mathrm{top}$ is independent of $\Delta H$ and only depends
on the basic parameters of that icy moon.

For realistic icy moons, $\kappa_v$ is highly uncertain because it may depend on complicated ocean tides \cite{Tyler-2011:tidal,chen2014tidal,idini2024resonant}, convection, and boundary layer processes \cite{large1994oceanic, lawrence2024ice}. The dependence of the freezing point on pressure
is constrained by the Clausius-Clapeyron equation, which depends only on the latent
heat of fusion and the density difference between water and ice. It gives a freezing point suppression rate of ${\mathrm{d}T_\mathrm{freezing}}/{\mathrm{d}p}
= 8\times 10^{-8}\mathrm{K}\,\mathrm{Pa}^{-1}$. 
If we further assume $\kappa_v = 10^{-4}\,\mathrm{m}^{2}\,\mathrm{s}^{-1}$ (not untypical of the Earth's ocean) and take $\rho_w = 1\times 10^{3}\,\mathrm{kg}\,\mathrm{m}^{-3}$
and $c_p = 4\times 10^{3}\,\mathrm{J}\,\mathrm{kg}^{-1}\,\mathrm{K}^{-1}$,
the upper bound is 0.5~GW for Enceladus, 235~GW for Europa,
and 690~GW for Titan assuming the parameters listed in Table~\ref{tab:icy-moons}.
Note that this upper bound only applies to positive $\Delta b$, which
could be violated by spatial variations in salinity and if $\alpha$ is negative.

This upper bound is reached when $\kappa_v$ approaches zero. In this limit the
isopycnals are all flat, so that there is no available potential energy to drive
the ocean circulation. None of our simulations reach this limit
(Fig.~\ref{fig:zonal-mean}). This limit is numerically challenging
because small $\kappa_v$ requires high spatial resolutions and long integration
times. In the limit of low diffusivity, our upper bound (Eq.~\ref{eq:OHT-bound})
can still provide a meaningful constraint on the oceanic meridional heat fluxes.

For negative $\alpha$, the upper bound discussed above is not valid. Instead, our scaling predicts a finite $|\psi^\star|$ (Eq.~\ref{eq:scaling-psi}) when $s =
s_\mathrm{top}$. Because the buoyant water source is below the dense water
source along the ocean top, the ocean system -- resembling rotating
Rayleigh-B\'enard along a titled surface -- still contains available potential
energy that can drive ocean circulation.

\subsection{Tidal heat production, ocean stratification, and ice shell's nonsynchronous rotation}


\begin{table}
    \caption{Parameters of Subsurface Oceans of Icy Moons}
    \centering
    \begin{tabular}{p{0.2\linewidth}p{0.2\linewidth}p{0.2\linewidth}p{0.2\linewidth}}
        \hline
        & Enceladus & Europa & Titan \\
        \hline
        $a$ (km) & 252 & 1561 & 2575 \\
        $\Omega$ ($\mathrm{s}^{-1}$) & $5.3\times 10^{-5}$ & $2.1\times 10^{-5}$ &
            $4.6\times 10^{-6}$ \\
        $g$ ($\mathrm{m}\,\mathrm{s}^{-2}$) & 0.1 & 1.3 & 1.4 \\
        $\kappa_\mathrm{mol}$ ($\mathrm{m}^2\,\mathrm{s}^{-1}$)
            & $1.4\times 10^{-7}$
            & $1.6\times 10^{-7}$
            & $1.8\times 10^{-7}$ \\
        $\rho_b$ ($10^3\,\,\mathrm{kg}\,\mathrm{m}^{-3}$) & 1.6 & 3.0 & 1.9 \\
        $D$ (km)
            & 40
            & 85
            & 369 \\
        $D_i$ (km)
            & 20
            & 15
            & 75 \\
        $T_s$ (K)
            & 59
            & 110
            & 94 \\
        \hline
        \multicolumn{4}{p{0.8\linewidth}}{note.
            Following \citeA{soderlund2019ocean}, we show
            the radius $a$, bulk density $\rho_b$,
            rotation rate $\Omega$,
            gravitational acceleration $g$,
            molecular diffusivity $\kappa_\mathrm{mol}$
            mean ocean depth $D$,
            mean ice thickness $D_i$,
            and surface temperature $T_s$
            for Enceladus, Europa, and Titan.
        }
        \\
    \end{tabular}
    \label{tab:icy-moons}
\end{table}

\begin{figure}
    \centering
    \includegraphics[width=0.8\linewidth]{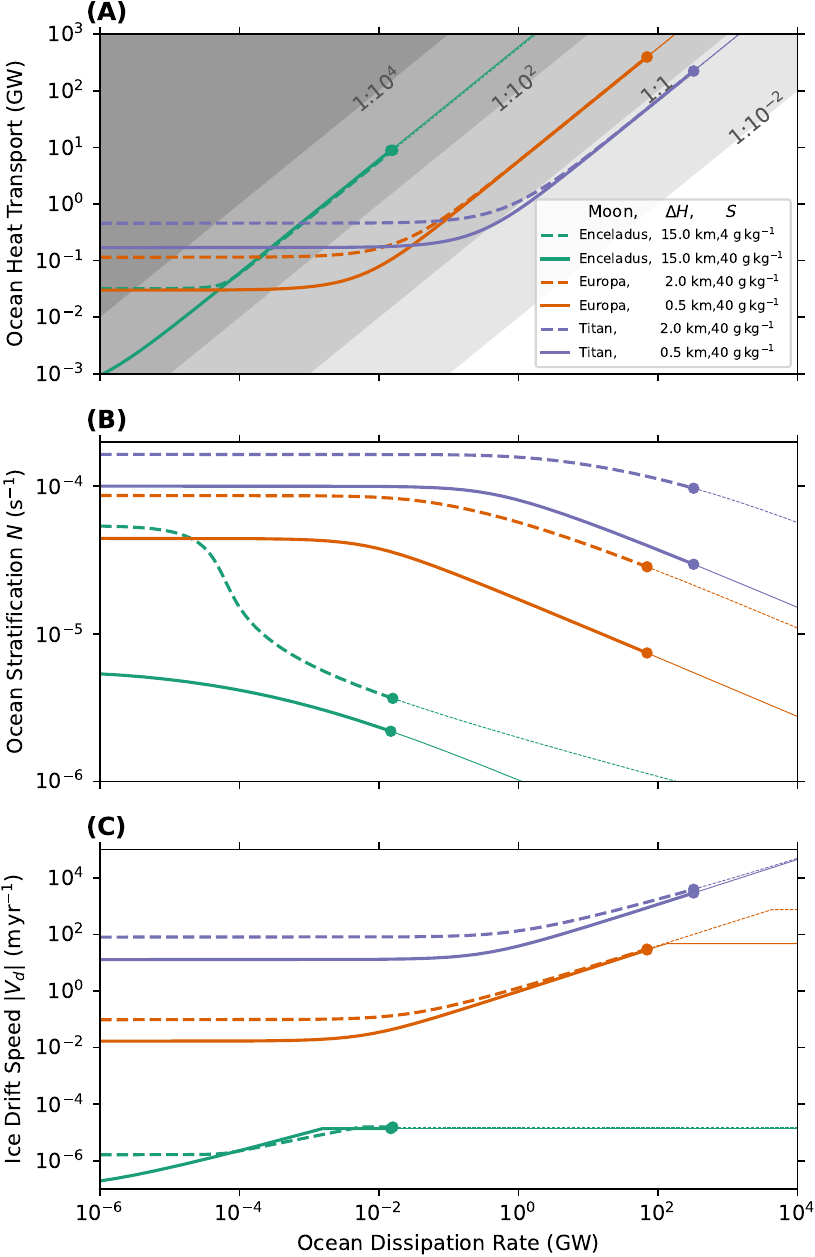}
    \caption{Implication for tidal dissipation, ocean heat transport, ocean
        stratification, and nonsynchronous rotation of the moon of large-scale ocean circulation. We plot (A) the meridional
        oceanic heat transport, (B) ocean stratification, and (C)
        nonsynchronous rotation rate of the ice shell driven by purported large-scale
        ocean circulation on Enceladus, Europa, and Titan for assumed
        equator-to-hole ice shell thickness variations $\Delta H$ and mean ocean
        salinities $S$. This calculation uses our scaling framework
        (Equations~\ref{eq:scaling-s-positive}, \ref{eq:scaling-s-negative},
        \ref{eq:scaling-psi}, and \ref{eq:scaling-f}) and Eq.~\ref{eq:kappa-v-scaling}.
        The parameters assumed for these three icy moons are
        given in Table~\ref{tab:icy-moons}.
        For each line, a dot marks the maximum allowed ocean
        dissipation that is consistent with the observed ice geometry (discussed
        in Sect.~\ref{sec:application}).
    }
    \label{fig:application}
\end{figure}

The vertical diffusivity $\kappa_v$ is a prescribed parameter in both our
numerical simulations and scaling analysis. In reality, $\kappa_v$ reflects a
combination of molecular diffusion and turbulent mixing driven by internal tide
breaking \cite{Osborn-1980:estimates} and convective plumes
\cite{lecoanet2013internal}. On Earth, $\kappa_v \sim
10^{-4}\,\mathrm{m^2\,s^{-1}}$ -- two to three orders of magnitude larger than
molecular diffusivity -- primarily due to tidal mixing
\cite{Munk-Wunsch-1998:abyssal,Wunsch-Ferrari-2004:vertical}.\

In our calculations, which lack bottom heating and convection, $\kappa_v$ can be expressed as
\begin{linenomath*}\begin{equation}
\kappa_v = \kappa_\mathrm{mol} + \Gamma \frac{\epsilon}{\rho_w N^2},
\label{eq:kappa-v-scaling}
\end{equation}\end{linenomath*}
where $\kappa_\mathrm{mol}$ is the molecular diffusivity, $\Gamma \sim 0.2$ is the mixing efficiency, $\epsilon = \dot{E}/V$ is the dissipation per unit volume, $\dot{E}$ is the total dissipation and $V$ is the ocean volume.

This allows us to reformulate our scaling framework using the total dissipation rate
$\dot{E}$ as the input parameter, substituting for $\kappa_v$ via
Eq.~\ref{eq:kappa-v-scaling}. This approach is advantageous because $\dot{E}$
can be directly estimated from tidal dissipation models for icy moon oceans.

Figure~\ref{fig:application}(A) shows the ocean heat transport (OHT) as a function
of $\dot{E}$. The ratio of OHT to $\dot{E}$ defines a thermal efficiency $\eta$,
which quantifies how effectively dissipated energy drives heat transport.
Following \citeA{jansen2022energetics}, $\eta \sim \Gamma c_p / (|\alpha|g
H_0)$, implying that Enceladus, with its low gravity, has a relatively high
efficiency, while Titan's stronger gravity leads to a lower $\eta$. When
$\dot{E}$ is large enough to dominate molecular diffusion, we find a
linear relationship between OHT and $\dot{E}$.

Since excessive equatorward OHT can cause net melting at low latitudes and erase
the poleward-thinning ice profile, it places an upper bound on the OHT and 
consequently on the allowable tidal dissipation rate $\dot{E}$. In
Fig.~\ref{fig:application}(A,B), the maximum allowed $\dot{E}$ is marked with
a point, and a thinner line style is used beyond this point to indicate
inconsistency with the observed or assumed ice geometry. The OHT threshold is
set equal to $\pi a^2 \mathcal{H}_\mathrm{cond}$
where $\pi a^2$ is $1/4$ of the surface area of the moon
and $\mathcal{H}_\mathrm{cond}$ is the estimated conductive heat flux through the
equatorial ice:
\begin{linenomath*}\begin{equation}
    \mathcal{H}_\mathrm{cond} = \frac{\kappa_0}{D_i}\log{
        \left(\frac{T_\mathrm{freezing}}{T_s}\right)
    }
    \label{eq:H-cond}
\end{equation}\end{linenomath*}
where $\kappa_0 = 651\,\mathrm{W}\,\mathrm{m}^{-1}$
\cite{Petrenko-Whitworth-1999:physics}; $T_\mathrm{freezing} \approx
273\,\mathrm{K}$; $D_i$ is the thickness of the ice shell; $T_s$ is the surface
temperature (listed in Table~\ref{tab:icy-moons}). Following Eq.~\ref{eq:H-cond},
the OHT threshold is
8.9~GW for Enceladus,
396~GW for Europa 
and 223~GW for Titan.


Furthermore, our scaling also predicts equilibrium ocean stratification
$N^2$ as a function of the tidal dissipation rate $\dot{E}$. As shown in
Fig.~\ref{fig:application}(B), $N^2$ decreases with increasing $\dot{E}$, since
stronger dissipation enhances mixing and drives the isopycnals farther apart,
thus weakening stratification.

In reality, there is a two-way feedback between $N^2$ and $\dot{E}$. On large
scales, ocean stratification is determined by the balance between baroclinic eddies and
vertical diffusion (which depends on $\dot{E}$), as captured by our model. However, on a small scale, the generation and dissipation of internal tides are
themselves sensitive to $N^2$
\cite{Tyler-2011:tidal, auclair2018oceanic, idini2024resonant}.
The intersection between our predicted curve $\dot{E}$-$N^2$ and those of
the tidal dissipation models determines the self-consistent equilibrium value of
$\dot{E}$.

Lastly, we predict the speed of nonsynchronous rotation (NSR) of the ice shell
driven by thermal wind, following \citeA{kang2024nonsynchronous}. When the
lateral buoyancy gradient near the ocean surface diffuses into the ocean
interior, it generates a zonal current (thermal wind), the strength of which
depends on the meridional buoyancy gradient and the depth over which it
penetrates. The resulting velocity difference across the vertical extent is
given by:
\begin{linenomath*}\begin{equation}
    \Delta_v U = \frac{\Delta b}{8a\Omega}\mathrm{min}\{as, D\}.
\end{equation}\end{linenomath*}
where $D$  is the depth of the ocean, which sets an upper bound on the depth of penetration. This zonal flow exerts a frictional stress on the overlying ice shell:
\begin{linenomath*}\begin{equation}
    \tau_s = (1/4)\rho_w C_d |\Delta_v U|\Delta_v U,
\end{equation}\end{linenomath*}
where $C_D$ is the drag coefficient, taken as $5\times 10^{-3}$ in this study.
This stress imparts a torque on the ice shell, leading to NSR at a drift speed
\cite{ashkenazy2023non}:
\begin{linenomath*}\begin{equation}
    V_d = k_0G^2\frac{\rho_b^2a^2}{\eta \Omega^4 D_i}
    \tau_s
    \left(1 - \frac{2\phi_\mathrm{TC} + \sin{(2\phi_\mathrm{TC})}}{\pi}\right)
\end{equation}\end{linenomath*}
where $k_0 = \frac{5\pi^3}{36}\frac{5 + \nu}{1 + \nu}\frac{1}{(1 + k_f)^2}$,
with the Poisson ratio $\nu = 1/3$ and the fluid Love number $k_f = 1.5$;
$G$ is the gravitational constant; $\rho_b$ is the bulk density
of the moon; the drag coefficient $C_d$ is assumed to be
$5\times 10^{-3}$; $\phi_\mathrm{TC}$ is the latitude at
which the tangent cylinder intersects with the ice;
$\eta$ is the depth-averaged viscosity of the ice shell, assumed
to be $1\times 10^{15}\,\mathrm{Pa}\,\mathrm{s}$ (\citeA{kang2024nonsynchronous}
contains a typo under Eq.~8: the default $\eta$ is $10^{15}\,\mathrm{Pa}\,\mathrm{s}$
rather than $10^{16}\,\mathrm{Pa}\,\mathrm{s}$).
Figure~\ref{fig:application}(C) shows the relationship between ocean
dissipation $\dot{E}$ and drift speed $|V_d|$.
As dissipation increases, the thermal wind penetrates deeper, strengthening the
zonal flow, and thus the NSR, until the penetration depth reaches the full ocean
depth. The drift speed $V_d$, which may be measurable by future missions, could
provide constraints on both oceanic circulation and tidal dissipation.

With our scaling framework, we unify several seemingly independent processes
through the lens of large-scale ocean circulation -- including ice shell geometry,
oceanic tidal dissipation, heat flux through the ice, and even the
non-synchronous rotation of the moon. This connection offers a pathway to better
constrain oceanic parameters and enables cross-validation between independent
observations. In the near future, Europa Clipper and JUICE (JUpiter ICy Moons
Explorer) will visit Europa, Ganymede, and Callisto, providing improved
measurements of non-synchronous rotation rates and ice thickness profiles. These
observations can, in turn, be translated into constraints on the oceanic
dissipation rate $\dot{E}$ and stratification $N^2$ using the analysis presented
here. Once $N^2$ is known, the tidal response of the ocean can be calculated from
first principles \cite{Tyler-2020:heating,
Rovira-Navarro-Rieutord-Gerkema-et-al-2019:do, Hay-Matsuyama-2019:nonlinear,
Rekier-Trinh-Triana-et-al-2019:internal} and potentially compared with future
seismic data, when available.

\section{Discussion and Conclusions}
    \label{sec:discussion}
    \label{sec:application}

This study investigates ocean circulation on icy moons driven by under-ice
temperature gradients arising from spatially varying ice shell thickness.
Motivated by the fact that tidal dissipation tends to be polar-amplified, we
consider a poleward-thinning ice shell following a cosine profile. The
Clausius-Clapeyron relation implies that the water beneath thicker equatorial ice is
colder than the water beneath the poles. Depending on the mean salinity and
pressure of the ocean, the thermal expansion coefficient $\alpha$ can be positive or
negative, determining whether the equatorial or polar water is denser.

Building on \citeA{zhang2024ocean} and \citeA{kang2022different}, we derive scaling laws for circulation strength and heat transport, incorporating the effects of ice topography and negative $\alpha$. We tested these predictions with 26 or so numerical simulations, exploring how topographic slope, vertical diffusivity, and buoyancy forcing influence ocean stratification and heat transport.

Our main conclusions are as follows.
\begin{enumerate}
    \item \textbf{Circulation structure}: Vertical diffusion energizes
        baroclinic eddies that can be represented as an overturning cell. This cell
        sinks at the equator when $\alpha>0$ and at the poles when $\alpha<0$.
        Regardless of the direction, heat flows from the poles (thin ice) toward the
        equator (thick ice), consistent with \citeA{kang2022does}.
    \item \textbf{Topographic suppression at positive $\alpha$}: When $\alpha >
        0$, the topography elevates the buoyant polar water above the denser
        equatorial water, weakening polar stratification and reducing meridional
        transport.
    \item \textbf{Topographic enhancement at negative $\alpha$}: When $\alpha <
        0$, topography lifts dense polar water above buoyant equatorial water,
 increasing the available potential energy \cite{jansen2022energetics}, thinning
        the stratified layer and enhancing heat transport. In this case, the heat flux
        remains finite even as $\kappa_v \to 0$.
    \item \textbf{Topography-dominant regime at small $\kappa_v$ and large $\Delta H$}:
        Topographic variations control ocean stratification and circulation
        when the topographic height difference exceeds the penetration depth set
        by vertical diffusion. In this regime, for positive $\Delta b$, the
        stratified layer conforms beneath the topography, with nearly flat
        isopycnals, leading to a weakened overall circulation. For negative
        $\Delta b$, the stratified layer forms a thin sheet that hugs the sloped
        topography. In both cases, the overturning circulation does not
        penetrate the ocean abyss.
\end{enumerate}

Finally, there are important caveats to our study. First, we do not account for salinity
variations, which can arise from melting and refreezing processes at the
ice-ocean interface \cite{ashkenazy2018dynamics,lobo2021pole,kang2022does}.
Second, we neglect bottom heating
\cite{melosh2004temperature, goodman2004hydrothermal, goodman2012numerical,
soderlund2014ocean, sekine2015high, soderlund2019ocean,
zeng2021ocean, bire2022exploring,
bire2023divergent, lemasquerier2023europa, bouffard2025seafloor},
which may influence ocean stratification and, through redistribution of heat,
also affect the geometry of the overlying ice shell. Finally, our simulations
are conducted in a Cartesian coordinate system, rather than a more realistic
spherical geometry.

%
%

%


%
%
%
%

\section*{Open Research Section}
Data used in this study will be available online upon acceptance of this
manuscript.

\section*{Conflict of Interest Statement}
The authors declare there are no conflicts of interest for this manuscript.

\acknowledgments
This project uses Oceananigans, an open source ocean general circulation model.
The authors thank Simone Silvestri, Gregory L. Wagner, and Xin Kai Lee for
technical support for Oceananigans and Yuchen Ma, Shuang Wang, Wenda Zhang, and
Kaushal Gianchandni for helpful discussions.
This work was supported in part by
NASA Astrobiology Grant 80NSSC19K1427 ``Exploring Ocean Worlds''.

%
%

\bibliography{citation}

%
%
%
%
%

\appendix

\section{Scaling for isopycnal slope}
    \label{appendix:isopycnal-slope-scaling}
    
At equilibrium, vertical heat diffusion and eddy heat
transport balance one-another:
\begin{linenomath*}\begin{equation}
    \partial_z\left(\kappa_v\partial_z\overline{b}\right)
        - \partial_y(\psi^\star \partial_z \overline{b})
        + \partial_z(\psi^\star \partial_y \overline{b}) = 0.
\end{equation}\end{linenomath*}
where $\psi^\star$ is the eddy-driven overturning streamfunction (shading
in Fig.~\ref{fig:zonal-mean}).

We consider its weak form:
\begin{linenomath*}\begin{equation}
    \int_\mathcal{D}
        \left(
            \partial_z\left(\kappa_v\left(\partial_z\overline{b}\right)\right)
            - \partial_y(\psi^\star \partial_z \overline{b})
            + \partial_z(\psi^\star \partial_y \overline{b})
        \right)\phi \, \mathrm{d}y\mathrm{d}z = 0,
\end{equation}\end{linenomath*}
where $\phi(y, z)$ is a test function; $\mathcal{D}$ is the
Northern Hemisphere of the ocean domain of $(y, z)$.

We perform integration by parts and use Green's theorem:
\begin{linenomath*}\begin{equation}
    \begin{aligned}
    &\int_{\partial\mathcal{D}}
        \left(
        (-\phi \psi^\star \partial_z \overline{b}) \mathrm{d}z
        + (\phi \psi^\star \partial_y \overline{b}
        + \kappa_v\phi \partial_z \overline{b}) \mathrm{d}y
    \right) \\
    + &\int_\mathcal{D} \left(
        - \kappa_v \partial_z \overline{b} \partial_z \phi
        + \psi^\star \partial_z \overline{b} \partial_y \phi
        -  \psi^\star \partial_y \overline{b} \partial_z \phi
    \right)  \mathrm{d}y\mathrm{d}z = 0.
    \end{aligned}
    \label{eq:full_weak_form}
\end{equation}\end{linenomath*}
where $\partial\mathcal{D}$ is the boundary of $\mathcal{D}$.

The contour integration term vanishes if we
choose $\phi = z - z_\mathrm{top}$.
Along the upper boundary, $\phi = 0$;
$\psi = 0$ at the equator and the North Pole;
$(-\kappa_v\partial_z\overline{b} - \psi^\star\partial_y\overline{b})$
is the net vertical buoyancy flux, which is zero as we do not consider
heating from below in this study.

For $\phi = z - z_\mathrm{top}$, $\partial_y \phi = -s_\mathrm{top}$
and $\partial_z \phi = 1$. Thus,
Eq.~\ref{eq:full_weak_form} becomes:
\begin{linenomath*}\begin{equation}
    \int_\mathcal{D} \partial_z \overline{b}\left(
        - \kappa_v
        + \psi^\star \left(\tilde{s} - s_\mathrm{top}\right)
    \right)  \mathrm{d}y\mathrm{d}z = 0,
\end{equation}\end{linenomath*}
where $\tilde{s} = -\partial_y\overline{b}/\partial_z\overline{b}$
is the signed isopycnal slope. The isopycnal slope we represent
in the main text is the absolute value $s = |\tilde{s}|$.
For positive $\Delta b$,
$\tilde{s} < 0$ and $\psi^\star < 0$ in the Northern Hemisphere;
For negative $\Delta b$,
$\tilde{s} > 0$ and $\psi^\star > 0$ instead. Taking into account the signs
of $\tilde{s}$ and $\psi^\star$, we get
\begin{linenomath*}\begin{equation}
    \int_\mathcal{D} \partial_z \overline{b}\left(
        - \kappa_v
        + |\psi^\star| \left(s + \mathrm{sgn}(\Delta b)s_\mathrm{top}\right)
    \right)  \mathrm{d}y\mathrm{d}z = 0,
\end{equation}\end{linenomath*}
which is a more rigorous form of 
Eq.~\ref{eq:scaling-kappa-over-psi}.

\section{Upper Bound on the Heat Flux for Positive $\alpha$}
    \label{appendix:upper-bound}

The upper bound of the heat flux can be derived from the buoyancy budget. The
steady-state buoyancy equation integrated over the domain gives the following:
\begin{linenomath*}\begin{equation}
    \partial_z\overline{wb} + \partial_y\overline{vb}
    - \partial_z(\kappa_v\partial_z \overline{b})
    - \partial_y(\kappa_h\partial_y \overline{b})
    = 0,
\end{equation}\end{linenomath*}
where buoyancy transport by Eulerian overturning
and horizontal diffusion is also included. Similarly to
\ref{appendix:isopycnal-slope-scaling}, we seek the weak form using $(z -
z_\mathrm{top})$ as the test function:
\begin{linenomath*}\begin{equation}
    \int_\mathcal{D} (z - z_\mathrm{top})\left(
    \partial_z\overline{wb} + \partial_y\overline{vb}
    - \partial_z(\kappa_v\partial_z \overline{b})
    - \partial_y(\kappa_h\partial_y \overline{b})
    \right)\mathrm{d}y\mathrm{d}z = 0
\end{equation}\end{linenomath*}
where $\mathcal{D}$ is the Northern Hemisphere where $y \in [0, (\pi/2)a]$
and $z \in [-D, z_\mathrm{top}(y)]$.

We integrate the equation by part. Using the no-flux boundary condition at the
meridional and bottom boundaries, the contour integration
term vanishes:
\begin{linenomath*}\begin{equation}
    \int_D \left(
    -\overline{wb} - s_\mathrm{top}\overline{vb}
    + \kappa_v\partial_z \overline{b}
    + s_\mathrm{top}\kappa_h\partial_y \overline{b}
    \right)\mathrm{d}y\mathrm{d}z = 0.
\end{equation}\end{linenomath*}

We separate the terms for the horizontal transport and get
a relation between the net horizontal transport weighted
by $s_\mathrm{top}$ and the vertical transport:
\begin{linenomath*}\begin{equation}
    \mathcal{F}_b
    = \frac{1}{\Delta H}\left(
    \int_\mathcal{D} \kappa_v \partial_z\overline{b} \mathrm{d}y\mathrm{d}z
    - \int_\mathcal{D} \overline{wb}\mathrm{d}y\mathrm{d}z
    \right)
\end{equation}\end{linenomath*}
where $\mathcal{F}_b$ is the vertical integrated meridional buoyancy transport
averaged with a weight of $s_\mathrm{top}$:
\begin{linenomath*}\begin{equation}
    \mathcal{F}_b = \frac{\int_\mathcal{D} s_\mathrm{top} \left(
        \overline{vb}
        - \kappa_h\partial_y \overline{b}
        \right)\mathrm{d}y\mathrm{d}z
    }{
        \int_{0}^{(\pi/2)a} s_\mathrm{top}\mathrm{d}y
    }.
    \label{eq:f-b-bound-d1}
\end{equation}\end{linenomath*}
Here, we have used $\int_{0}^{(\pi/2)a} s_\mathrm{top}\mathrm{d}y = \Delta H$.

The vertical diffusion term is bounded by:
\begin{linenomath*}\begin{equation}
    \int_\mathcal{D} \kappa_v \partial_z\overline{b} \mathrm{d}y\mathrm{d}z
    \le \kappa_v ((\pi/2)a) |\Delta b|
    \label{eq:f-b-bound-d2}
\end{equation}\end{linenomath*}
where $(\pi/2)a$ is the horizontal extent of $\mathcal{D}$.

The vertical buoyancy flux $\overline{wb}$, the rate at which the potential energy
is released and drives the circulation, must be nonnegative:
\begin{linenomath*}\begin{equation}
    \int_D \overline{wb}\mathrm{d}y\mathrm{d}z \le 0
    \label{eq:f-b-bound-d3}
\end{equation}\end{linenomath*}

Combing Equations~\ref{eq:f-b-bound-d1}, \ref{eq:f-b-bound-d2}, and
\ref{eq:f-b-bound-d3}, we get:
\begin{linenomath*}\begin{equation}
    \mathcal{F}_b \le \frac{(\pi/2)a}{\Delta H}\kappa_v |\Delta b|,
\end{equation}\end{linenomath*}
where is a more rigorous form of Eq.~\ref{eq:f-b-bound}.

\end{document}